\documentclass{article}

\usepackage{arxiv}

\usepackage[utf8]{inputenc} 
\usepackage[T1]{fontenc}    
\usepackage{hyperref}       
\usepackage{url}            
\usepackage{booktabs}       
\usepackage{amsfonts}       
\usepackage{nicefrac}       
\usepackage{microtype}      
\usepackage{lipsum}		
\usepackage{graphicx}
\usepackage{natbib}
\usepackage{doi}
\usepackage{multirow}
\usepackage{amsmath}
\usepackage{listings}
\usepackage{float}
\usepackage{xcolor}

\title{From 2G to 4G Mobile Network: Architecture and Key performance indicators }

\author{ \href{https://orcid.org/0000-0002-9532-2453}{\includegraphics[scale=0.06]{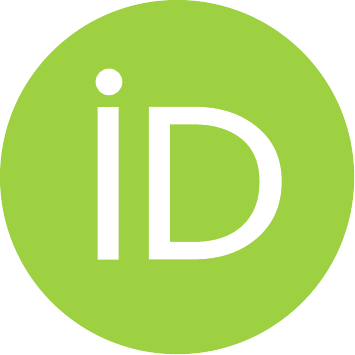}\hspace{1mm}Hamza Kheddar}\thanks{Corresponding author} \\
	Department of Electrical engineering, 
	University of Medea\\
	Medea, 26000, Algeria \\
	\texttt{kheddar.hamza@univ-medea.dz} \\
}	

\renewcommand{\shorttitle}{From 2G to 4G Mobile Network: Architecture and Key performance indicators}

\hypersetup{
pdftitle={A template for the arxiv style},
pdfsubject={q-bio.NC, q-bio.QM},
pdfauthor={Hamza Kheddar},
pdfkeywords={First keyword, Second keyword, More},
}

\begin{document}
\maketitle

\begin{abstract}
The second-generation (2G) mobile systems were developed in response to the growing demand for a system that met mobile communication demands while also providing greater interoperability with other systems. International organizations were crucial in the development of a system that would offer better services, be more transparent, and be more interoperable with other networks. The aim of having a single set of standards for networks worldwide was sadly not realized by the 2G network standards. The third generation (3G) was born. It was called the universal terrestrial mobile system (UMTS), which is European telecommunications standards institute (ETSI) driven. IMT‐2000 is the international telecommunication union-telecommunication standardization sector (ITU‐T) name for the 3G network. Wide-band code division multiple access (WCDMA) is the air interface technology for the UMTS. This platform offers many services that are based on the Internet, along with video calling, imaging, etc.
Further advancements to mobile network technology led to long term evolution (LTE), a technology referred to as 4G. The primary goal of LTE was to improve the speed and capacity of mobile networks while lowering latency. As we move to an ALL-IP system, mobile networks' design becomes much simpler. LTE uses orthogonal frequency division multiplexing (OFDM) in its air interface. This paper details all mentioned mobile generations, as well as all the differences between them in terms of hardware and software architectures.

\end{abstract}


\section{Introduction}
\label{sec:1}
Mobile network evolution has been categorized into \textit{generations} as a differentiating factor between several technologies and their capabilities. It all started with the first generation (1G) in the 1980s, which was based on analogue transmission techniques. There was no international coordination for the creation of the system's technical standards at the time. Nordic countries have deployed Nordic mobile phones (NMTs), while the United Kingdom and Ireland chose a complete access communication system (TACS). They primarily provided speech services and were incompatible with each other. As a result, their principal constraints were the restricted number of services available and their incompatibility. The creation of second-generation (2G) mobile systems stemmed from an increased requirement for a system that catered to mobile communication needs while also offering enhanced interoperability with other systems. International organizations were critical in developing a system that would deliver better services while also being more transparent and interoperable with networks worldwide. However, the 2G network standards were unable to realize the aim of having a single set of standards for networks worldwide. For example, the standards in Europe differed from those in Japan and the Americas \cite{mishra2018fundamentals}. Among all the standards, the global system for mobile communication (GSM) fulfills the technical and commercial expectations. GSM experienced changes in the form of general packet radio services (GPRS) and enhanced data rates in GSM environment (EDGE).
EDGE allowed for very high-speed data transportation, but the packet transmission on the air interface still behaved as a circuit switch call. As a result, in the circuit switch scenario, some of the packet connection efficiency is lost. Furthermore, in 2G, the criteria for establishing networks varied in different regions of the world. As a result, it was determined to build a network that offers services independent of the technological platform and has worldwide network design standards \cite{mishra2018fundamentals}. Thus, 3G was born, it was called UMTS, which is ETSI-driven. IMT‐2000 is the ITU-T standard name for 3G, while CDMA2000 is the name of the American 3G variant. Wide-band code division multiple access (WCDMA) is the air interface technology for UMTS. This platform offers many services that are based on the Internet, along with video calling, imaging, etc. This platform provides several Internet-based services, such as video calling, image sharing, and so on. Further improvements in mobile network technology resulted in long-term evolution (LTE), sometimes known as 4G. The primary goal of LTE was to enhance the speed and capacity of mobile networks while decreasing latency. As the public land mobile network (PLMN) go toward an ALL IP system, the design of mobile networks becomes easier. LTE employs orthogonal frequency division multiplexing (OFDM) as an access technique in its air interface \cite{mishra2018fundamentals}.

In this paper, the author introduce the architecture and the protocols involved in building 2G, 3G, and 4G mobile networks, with the goal of further understanding the optimization process for a satisfactory experience for all users.

\section{The second generation mobile network}
The 2G of digital mobile phones, as well as the first digital mobile networks, arrived in the 1990s. The mobile telecommunications sector grew exponentially in terms of customers and value-added services (VAS) throughout the second generation. Limited data support is available on 2G networks, rising from 9.6 kbps to 19.2 kbps \cite{mishra2007advanced}.
From 1982 to 1985, members of the global system for mobile (GSM) group, which was initially hosted by the European conference of postal and telecommunications administrations (CEPT), debated whether to construct an analog or digital system. It was determined to develop a digital system after several field tests. A narrow-band time division multiple access (TDMA) solution was chosen, and Gaussian minimum shift keying (GMSK) has been selected as modulation strategy. By 1987, the technological foundations had been established, and the first specification had been written by 1990. With Radiolinja in Finland, GSM became the first commercially functioning digital cellular system in 1991. With over a billion people utilizing the system, GSM was the most popular and commonly used cellular system in 2005. The system's popularity was enhanced by features like prepaid calling, international roaming, and so on. Naturally, this resulted in the creation of smaller, lighter handsets with a greater number of functions. Apart from making phone calls, the system grew more user-friendly by adding a variety of services. Voice mail, short message service (SMS), call waiting (CW), and other services were available. By the year 2000, about 15 billion SMS had been sent every month, indicating that SMS had been a huge success. The higher quality of digital voice and low-cost options to communicate, such as text messaging, have been the primary advantages of GSM networks. Because the open standard enables easy interoperability, network operators have been able to install equipment from many providers.
In this section, an overview of the GSM network’s architecture with its different subsystems and the functions of each one, the different processes that happen within the network such as handover and roaming, all the iterations of 2G such as GPRS, and EDGE. The upgrades that were made to the architecture and, most importantly, the process of the network optimization are thoroughly taken into consideration.

\subsection{GSM communication system}
\label{sec:2}
The GSM standard was invented by ETSI to outline the 2G digital network protocols employed by terminals such as mobile phones. By the mid-2010s, it had become a worldwide standard for mobile communication , spreading over more than 90\% and established in 193 countries' operators \cite{mishra2007advanced}.

\subsubsection{GSM Architecture }
\label{sec:3}

The main 2G PLMN subsystems are: 

\begin{itemize}
\item \textbf{The mobile station (MS)}: There are specific entities that used by mobile users to access many services. This latter are made up of a number of important components:
    \begin{itemize}
        \item \textbf{Mobile equipment (ME)}: Or, as they're more often known, cell phones, which are the only GSM terminals that  enables users to access the GSM network.
        \item \textbf{Subscriber identity module (SIM)}:  gives a unique identity to mobile equipment. Certain subscriber characteristics, as well as the subscriber's personal data, are saved on the SIM card. The subscribers are identified by the SIM card in the network. Before using the phone, customers must enter a 4-bit personal identification number (PIN) to secure the SIM card from unauthorized usage.
    \item \textbf{International mobile station equipment identity (IMEI)}: It is a unique serial number used to identify mobile stations all around the world. The equipment manufacturer assigns it, and network operators record it in the equipment identification registry (EIR). The IMEI is a hierarchical address that consists of the following components: 
    (i) Type approval code (TAC): Six digits, it is typically associated with the cellular modem or module.  (ii) Final assembly code (FAC): Identifies the location of the last production stage and consists of two digits. (iii) Serial number (SNR): Consists of six digits, assigned by the manufacturer. (iv) Spare (SP): Consist of one digit. Figure \ref{imei} depicts the main components of the IMEI number.

\begin{figure}
	\centering
	\includegraphics[scale=0.6]{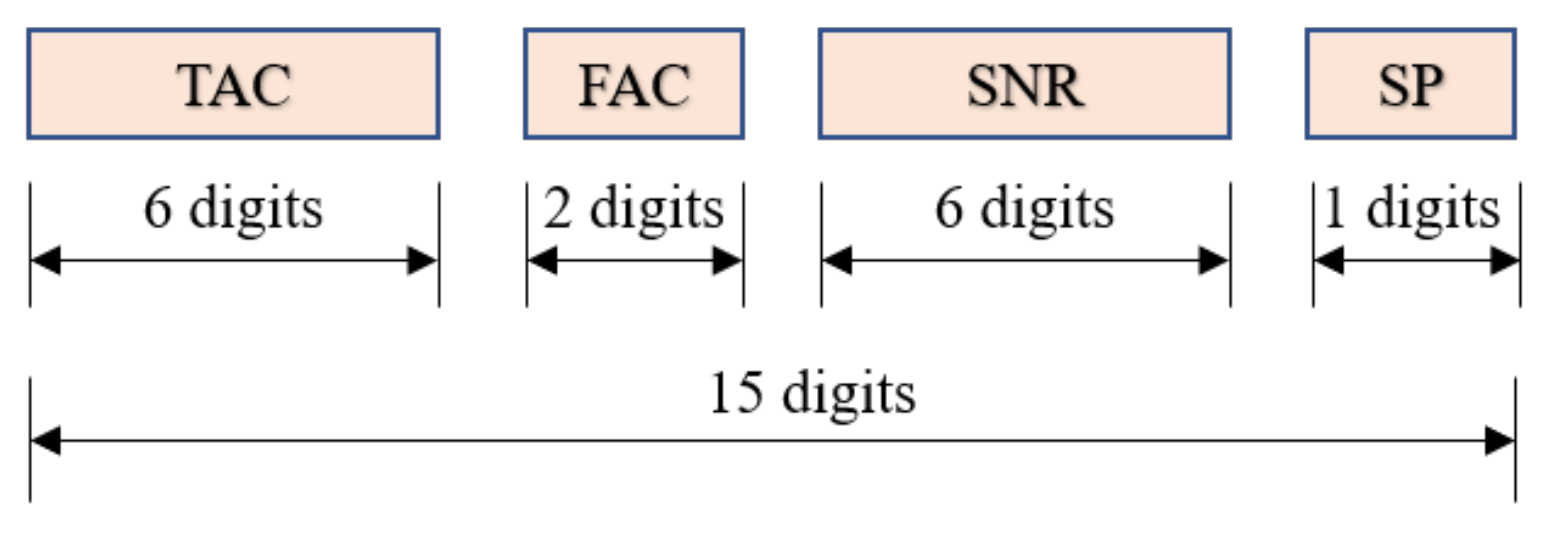}
	\caption{Structure of the IMEI number.}
	\label{imei}
\end{figure}

\item \textbf{International mobile subscriber identity (IMSI)}: Each subscriber obtains a unique identification called IMSI upon registering for service with a network operator, which is registered in the SIM. The mobile equipment with a valid IMEI works properly if the SIM card with a valid IMSI is plugged in it.
\item \textbf{Mobile subscriber ISDN number (MSISDN)}: The MSISDN is the MS's actual phone number. It is assigned to the subscriber; hence one MS is able to have many MSISDNs based on the SIM card. The MSISDN and IMSI association registered in the home location registry (HLR) must be known before the subscriber identification can be derived from the MSISDN.
\item \textbf{Mobile station roaming number (MSRN)}: Is a temporary integrated service digital network (ISDN) number that changes based on the location. It is assigned to each MS in its territory by visitor location registry (VLR), the MSRN is used to route MS calls. On request, the MSRN is sent from the HLR to the gateway mobile switching (GMSC).
\end{itemize}

\item \textbf{The base station subsystem (BSS)}: Is primarily responsible for connecting the mobile devices to the network. It is made up of two parts:

\begin{itemize}
    \item \textbf{Base transceiver station (BTS)}: The radio transmitter receivers, as well as their accompanying antennas, are part of the BTS in a GSM network, they broadcast and receives signal to connect to the mobile directly. Each cell is defined in a specific BTS. The BTS exchange communication with the mobile phones through the Um interface, which has its own set of protocols.
   \item \textbf{Base station controller (BSC)}: Controls the radio subsystem, specifically the BTSs. The main roles of the BSC are the radio resources and handover management. It is also responsible for managing the power control. It controls the signaling, and the security of the operation and maintenance (OM) subsystem. The BSC uses the Abis interface to connect with the BTSs \cite{pattem2022research}.
\end{itemize}

\item \textbf{Network switching subsystem (NSS)}:  Handles call switching between mobile and other fixed or mobile network users, as well as mobile service administration such as authentication. The NSS system includes the following functional elements:
\begin{itemize}
    \item \textbf{Mobile switching center (MSC)}: Is the most important component of the entire GSM network architecture's core network area. It functions similarly to a standard switching node in a public switched telephone network (PSTN) or an ISDN, but adds capabilities to satisfy the needs of mobile users such as authentication call location, call routing and subscriber location, inter-MSC handover, and registration. It also plays the role of a gateway to the PSTN, enabling mobile communications calls to be routed from the PLMN to a PSTN. It connects to other MSCs, allowing calls to be established to mobile phones on various operators. 
    \item \textbf{Home location register (HLR)}: It's a database that stores all the subscriber’s administrative information as well as their last known location. The MSC may then route calls to the appropriate BTS where the MS is located. When a user turns on their mobile, it registers with the network, allowing the BTS that it connects, to be identified and the incoming calls is then routed correctly to the appropriate subscriber. Even when the phone is turned off (or idle mode), it registers the new state to guarantee that the HLR is up to date on its location. Although it may be deployed over several sub-centers for operational reasons, each network can has only one HLR \cite{mishra2007advanced}. 
    \item \textbf{Visitor location register (VLR)}: It comprises information from the HLR that permits and provides the visitor subscriber’s services. Although the VLR can be designed as a distinct entity, it is more implemented done as an integrated element of the MSC, this improves access by making it faster and more efficient. 
    \item \textbf{Equipment identity register (EIR)}: Is the authority that decides whether to allow certain mobile devices into the network. Every mobile device has a number called IMEI. As described previously, this number is installed in the device and checked by the network during registration. Based on the information stored in the EIR, the phone can be assigned one of three states - allowing access to the network, blocking access, or monitoring if something goes wrong. 
    \item \textbf{Authentication center (AuC)}: Is a database that stores the secret key, which may also be found on the user's SIM card. On the radio channel, it's utilized for ciphering and authentication.
    \item \textbf{Gateway mobile switching center (GMSC)}: Is the first point to which a ME terminating call is routed, even if the MS's location is unknown. As a result, the GMSC is responsible for getting the MSRN from the HLR based on the MSISDN and routing the call to the proper visiting MSC.  
\end{itemize}

\item \textbf{Operation and support subsystem (OSS)}: Is a supervision component of the overall GSM mobile communications network architecture, it is linked to NSS and BSS. It is utilized to regulate and monitor the whole GSM network, as well as supervise the BSS's traffic load. The three major functions of OSS are:
\begin{itemize}
    \item Maintain all telecommunication hardware and network operations with a particular market.
    \item Manage all charging and billing procedures.
    \item Manage all mobile equipment in the system.
\end{itemize}
\end{itemize}

\begin{figure}
	\centering
	\includegraphics[]{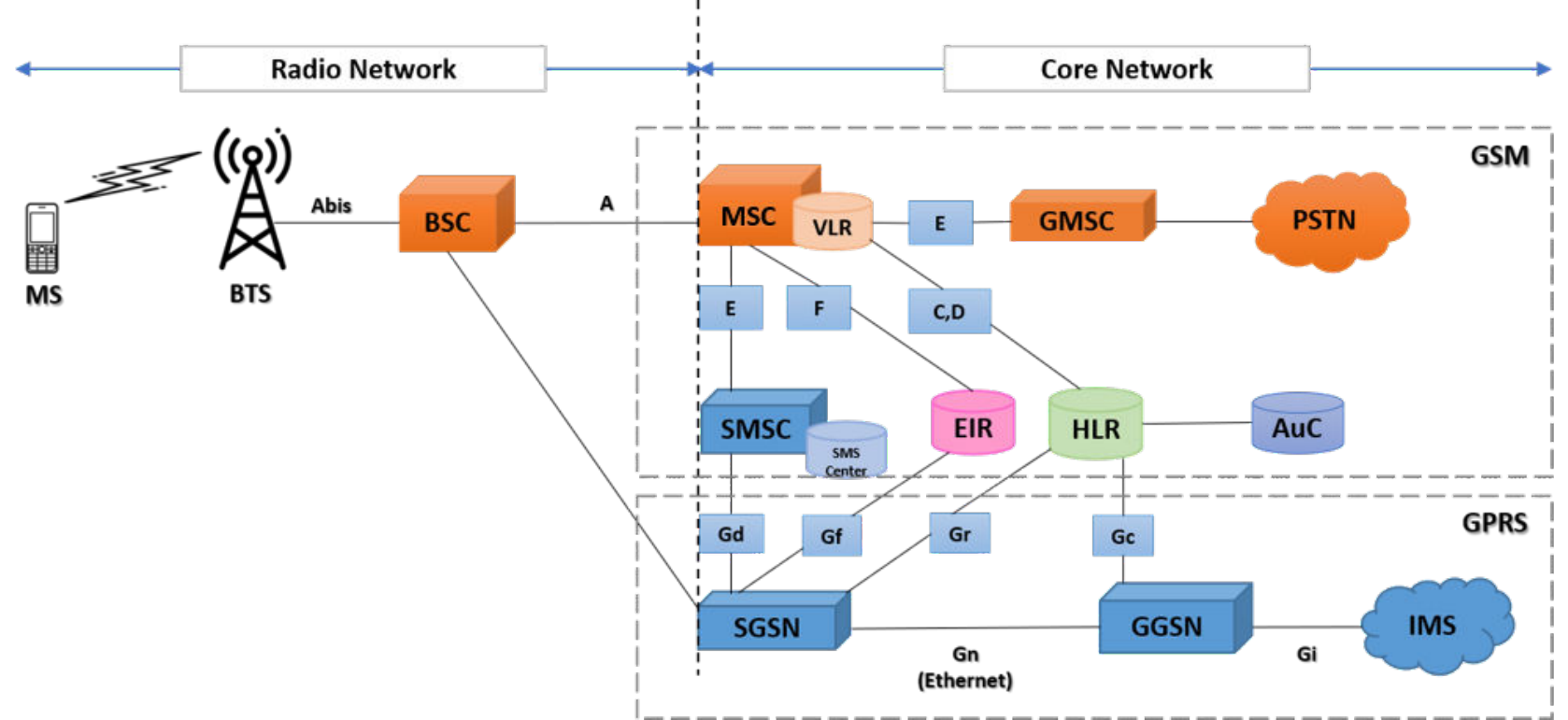}
	\caption{GSM globale architecture.}
	\label{figgsm}
\end{figure}

Figure \ref{figgsm} depicts the architecture of GSM and its entities.

\subsubsection{GSM Interfaces }
Standard interfaces were created to guarantee that networks could be constructed using components from various manufacturers. This ensured that they communicated in the same way regardless of the manufacturers. Table \ref{gsmInterfaces} shows the interfaces in the GSM network.

\begin{table}[]
	\caption{GSM network interfaces.}
	\label{gsmInterfaces}
\begin{tabular}{p{2.5cm}p{3.5cm}p{9cm}}
\hline
\textbf{GSM interface} & \textbf{Link}  & \textbf{Role} \\
 \hline
Um & MS and BTS & Carries the GSM bursts carrying data and control information. Also referred as Air interface. \\

Abis & BTS and BSC & Supports two types of communication links via traffic channel at 64 kbps and signaling channel at 16 kbps. \\

A & BSC and MSC/VLR & Supports 2Mbps standard digital connection as per CCITT. \\

 B & MSC and VLR & Used whenever the MSC needs access to data regarding a MS located in its area.\\

 C & HLR and GMSC.
Also, between MSC and HLR. & Used to obtain the requiring information required to complete the call.\\

 D & HLR and VLR & Used to exchange the data related to the location of the ME and to the management of the subscriber.\\

 E & MSC and another MSC or G-MSC. & Exchanges data related to handover between the anchor and relay MSCs.\\
 
F & EIR and MSC and between EIR and G-MSC.& Used to confirm the status of the IMEI of the ME gaining access to the network.\\
 
 G& VLR and another VLR. & Used to transfer subscriber information.
\\
 \hline
\end{tabular}
\end{table}

\subsection{The handover in GSM}

\begin{figure}
	\centering
	\includegraphics[scale=0.8]{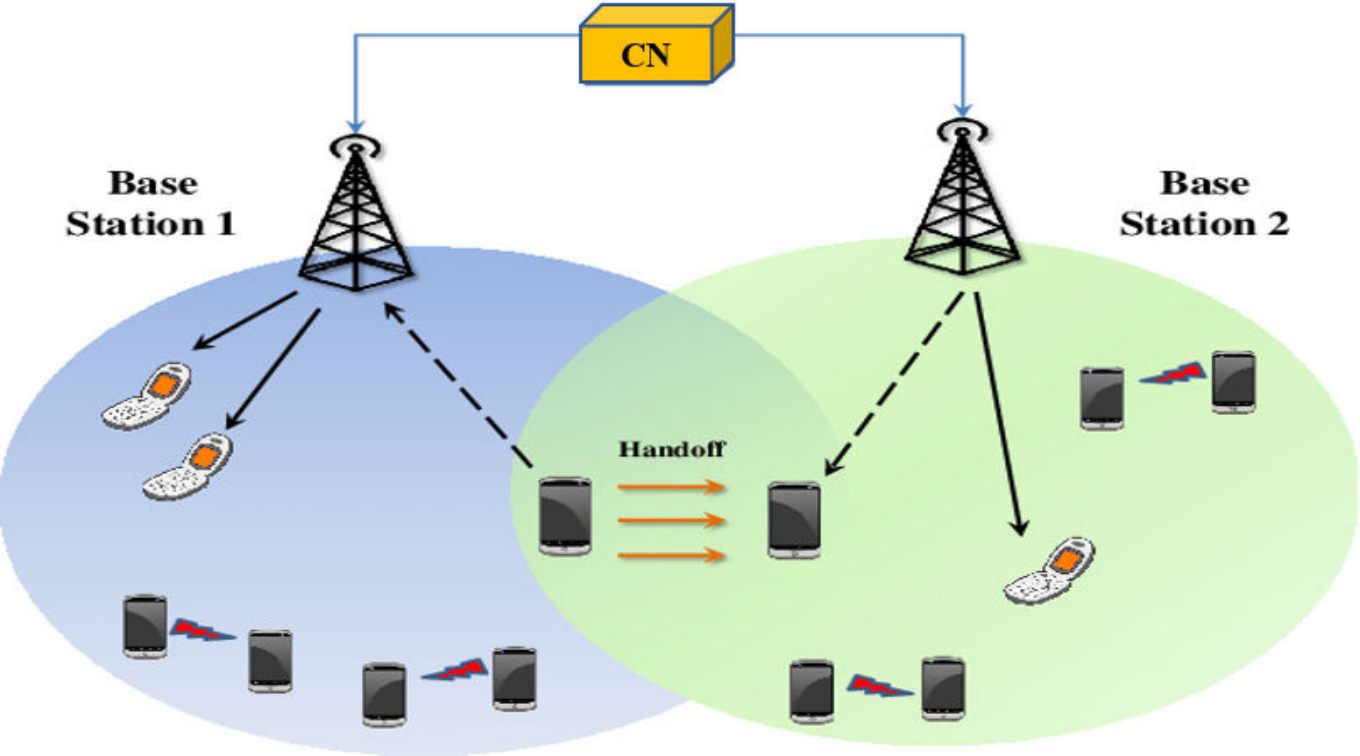}
	\caption{Handover in GSM network.}
	\label{figHOGSM}
\end{figure}

One of the most important features of a mobile phone or cellular telecommunications system
is that it is divided into numerous small cells in order to maximize frequency reuse and coverage. However, it must be feasible to maintain the connection while the phone goes from one cell to the next. The procedure is called as handover or hand-off. The handover process within any cellular system is very important. This is a critical process that can result in lost calls if not switched correctly. Calls drop are especially annoying to users, and when the number of calls drop rise up, customer dissatisfaction increases and they may probably to switch to another network. Therefore, GSM handover is a field of special concern in the standard development process. Figure \ref{figHOGSM} shows the steps of handover process. When one subscriber moves from blue cell to the green cell, the handover process launched to maintain the call.

\subsubsection{Different types of handover}
\label{sec:ho}

For GSM-only systems, there are four forms of handover that may be conducted inside the GSM system: 

\begin{itemize}
    \item \textbf{Intra-BTS handover}: This type of handover happens when a mobile phone's frequency or slot needs to be changed due to interference. In this type of GSM handover, the mobile phone stays connected to the same BTS, but the channel or slot is switched.
    \item \textbf{Inter-BTS intra BSC handover}: It happens when a mobile phone leaves the area of one BTS and goes to another managed by the same BSC. Before releasing the link between the former BTS and the MS, the BSC can launch the handover and allocate a new slot in a new channel.
    \item \textbf{Inter-BSC handover}: It happens when the mobile leaves the range of cells mastered by one BSC, the MS passed from one BTS to another BTS that belong to another BSC. The MSC is in charge of this handover in this case.
    \item \textbf{Inter-MSC handover}: This type of handover occurs when mobile leave MSC area to another MSC area, the two concerned MSC negotiate to master the handover.
\end{itemize}
\subsubsection{Roaming}
\label{roming}

Roaming permits mobile users to automatically send and receive data, attempt to send and
receive calls, or exploit other GSM services while traveling outside the geographic coverage
area of their network via a visiting network. Authentication, billing, and mobility management
procedures support roaming. This latter, called domestic, when visiting and home networks are in the same country, otherwise, it called international roaming. The inter-standard roaming
happens when the visiting and home networks use different technical standards. Roaming in GSM, allows mobile subscribers to utilize a single number, a single bill, and a single phone in as many as 219 countries.

\subsection{GSM radio transmission }
 
\subsubsection{Frequency bands in GSM}

GSM network occupies two frequency bands at around 900 MHz, which are:
\begin{itemize}
    \item \textbf{Up-link}: From 890 to 915 MHz.
    \item \textbf{Down-link}: From 935 to 960 MHz.
\end{itemize}

Each bandwidth is divided to 125 channels (each channel is 200 KHz). These channels are not sufficient in urban areas, so it’s necessary to attribute an additional band (DCS 1800) which has the same characteristics as the GSM 900 band in terms of protocols and services. DCS
occupies two frequency bands around 1800 MHz, which are:
\begin{itemize}
    \item \textbf{DCS Up-link}:  From 1710 to 1785 MHz.
    \item \textbf{DCS Down-link}: From 1805 to 1880 MHz.
\end{itemize}

Table \ref{gsmdcsCaract} summarizes most of signal characteristics of GSM-900 and GSM-1800 for both up-link and down-link.

\begin{table}[]
\centering
	\caption{Characteristics technique of GSM Network.}
	\label{gsmdcsCaract}
\begin{tabular}{p{6cm}p{4cm}p{4cm}}
\hline
 & \textbf{GSM-900}  & \textbf{DCS-1800} \\
 \hline
Downlink (MHz) & 935 to 960 & 1805 to 1880 \\

Uplink (MHz) & 890 to 915 & 1710 to 1780\\

Links bandwidth (MHz) & 45 & 95 \\

Number of channels & 125 & 375\\

Channels bandwidth (KHz) & 200 & 200\\

Number of time slots (TDMA) & 8 & 8\\

Logical channels number & 992 & 2992\\
 \hline
\end{tabular}
\end{table}

\subsubsection{Radio access techniques in GSM}
GSM networks use digital technologies to transmit data through radio waves. Frequency division multiple access (FDMA) and time division multiple access (TDMA) are the two technologies used in GSM.

\begin{itemize}

\item \textbf{TDMA}: Is a communications technique that allocates distinct time slots to various users on a shared channel (multi-points of broadcast) enabling conversations among several
users. Figure \ref{fdmaTdma} (a) illustrates the principle of TDMA.

\item \textbf{FDMA}: Is a technique for extracting, from transmission bearer, two or more continuous and simultaneous channels by allocating a different piece of the available frequency spectrum to each channel. It's used to send several broadcasts to a single transponder at the same time. The term FDMA refers to the fact that each BTS has its own radio frequency (RF) channel. MS in neighboring cells (or the same cell) can work at the same time, although they are separated by frequency. Multiple carrier frequencies, 125 in GSM 900 and 375 in GSM 1800, are used in the FDMA technology. Figure \ref{fdmaTdma} (b) illustrates the principle of FDMA.

\begin{figure}
	\centering
	\includegraphics[scale=0.9]{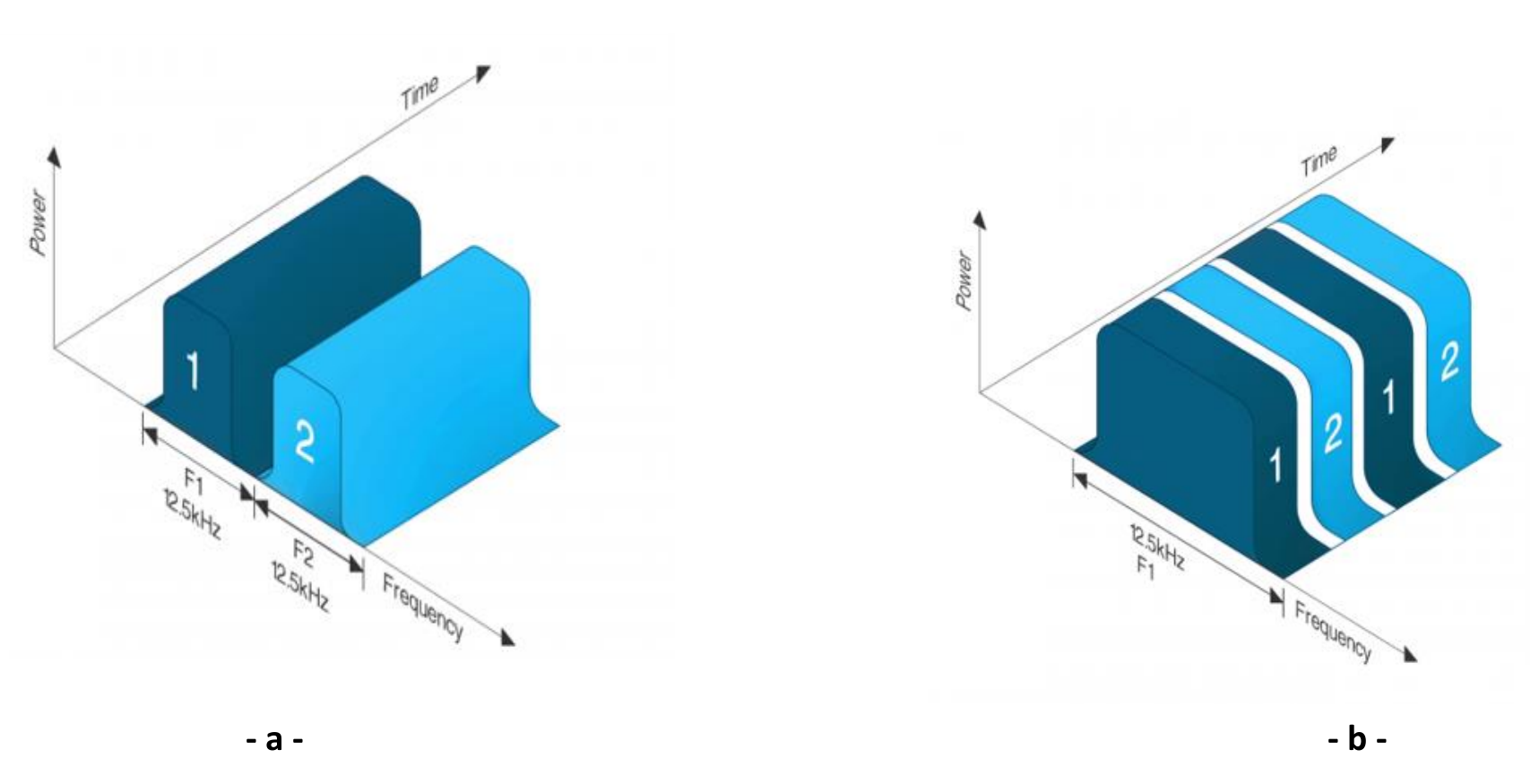}
	\caption{Time/frequency diagram of Access techniques. (a) TDMA, (b) FDMA.}
	\label{fdmaTdma}
\end{figure}

\end{itemize}

\subsection{Logical channels in GSM}
In GSM, logical channels are used to transport traffic and control data. The traffic channels are used to send information to users (speech or data). Control and signaling information are
transmitted over the control channels (CCHs), and used for short-message services (SMS). The cell broadcast channel (CBCH) is used to send user information from a service center to mobile stations in a particular cell area.
Common control channel (CCCH), dedicated control channel (DCCH), and broadcast
channel (BCH) are the control channels. The BCHs are used for frame synchronization, frequency correction, and control channel structure. They are down-link-only with point-to-multi-point channels. The broadcast control channel (BCCH), frequency correction channel
(FCCH), and synchronization channel (SCH) are among these channels. The BCCH is used to send cell identities, information about CCCHs, available cell service,
and so on. Frequency correction data bursts containing all "0" are transmitted over the frequency correction channel. This results in a constant frequency shift in the RF carrier, which the mobile station can be used for frequency correction. The SCH is employed to keep mobile stations' time synchronized. Frame number and the BTS identity code (BSIC), which is required by MSs when measuring base station signal strength, are among the data on this channel. Paging channel (PCH), random-access channel (RACH), and access grant channel (AGCH) are examples of CCCHs. The mobile stations are paged using PCHs. At certain times, the mobile stations must listen for paging. The AGCHs are used to assign mobiles to stand-alone dedicated control channels (SDCCHs). The RACHs are uplink-only channels that MSs use to
transmit requests for dedicated network connections.

The SDCCH and the associated control channel (ACCH) are the two types of DCCHs. SDCCHs are point-to-point bidirectional channels that are used for ciphering initiation, service requests, subscriber authentication, traffic channel (TCH) assignment, and equipment validation. The
ACCHs are point-to-point bidirectional channels that are linked to a specific TCH and SDCCH. Out-of-band signaling and control data are sent between the MS and the BTS over these channels. Signal strength measurements from a MS to the BTS and transmission timing information from the BTS to the MS are two examples of their use. Table \ref{logicalChan} summarized most of the existing logical channels and their roles in GSM network \cite{garg2010wireless}.

\begin{table}[]
\centering
	\caption{Most of the logical channels in the GSM system.}
	\label{logicalChan}
\begin{tabular}{p{3cm}p{12cm}}
\hline
\textbf{Logical channel} & \textbf{Role of the logical channel}   \\
 \hline
TCH/F &  Full rate traffic channel to carry payload at 22.8 kbps.\\
TCH/H & Half rate traffic channel to carry payload at 11.4 kbps.\\
BCCH &  Broadcast network information. \\

SCH & Synchronization of the mobile stations.\\

FCCH & Used for frequency correction. \\

AGCH & Accepts a channel request from a mobile device and
assigns an SDCCH to it.\\

FACCH & Traffic burst is stolen for a full signaling burst TCH inband signaling, e.g., for link monitoring, for time critical
signaling over TCH (e.g., for handoff signaling).\\

SACCH & TCH in-band signaling, e.g., for link monitoring.\\

SDCCH & Registration/location updates FACCH for SDCCH during
signaling exchange, such as during call setup.
For a full signaling burst, the SDCCH burst is stolen.\\

FACCHs& For SDCCH, FACCH is the acronym. For a full signaling
burst, the SDCCH burst is stolen.\\

SACCHs & SDCCH in-band signaling, e.g., for link monitoring.\\
 \hline
\end{tabular}
\end{table}

\subsection{The GPRS}

For 2G cellular systems, GPRS provided a packet data capability, allowing GSM to evolve
to include a data capability. Additional network entities must be added to the overall
architecture for the latter purpose; three of the most important are the GPRS gateway service
node (GGSN) and the serving GPRS support node (SGSN), and  the packet control unit (PCU) as shown in Figure \ref{figgsm}. GPRS to provide additional connectivity in terms of packet data.

\begin{itemize}
\item \textbf{SGSN}: Provides a variety of services focused on the IP parts of the overall system. It offers a wide range of mobile services, including: Packet routing and transfer, mobility management, authentication, charging data.
\item \textbf{GGSN}: Is a critical component of the GPRS network design. It can be thought of as a hybrid of a gateway, router, and firewall. When the GGSN receives data addressed to a specific user, it first verifies that the user is online before sending the data.
\item \textbf{PCU}: Which is connected to the BSC, is a hardware router. It converts data for the standard GSM network (circuit switched data) to data for the GPRS network (GPRS data) and vice-versa.
\end{itemize}

\subsection{The EDGE}
With the introduction of EDGE technology, some of the limitations of the GPRS network were alleviated. Both TDMA and GSM networks are compatible with EDGE. Because it can be installed on any machine that supports GPRS, it is considered a subset of GPRS. The theoretical data rate for EDGE is 500 kbps. The phase-shift keying (PSK) modulation technique is used to achieve the theoretical throughput. EDGE features increase the throughput per time slot (8.8–59.2 kbps/time slot), modulation changes from GMSK in GPRS to 8-PSK, decrease the signal sensitivity, and increase capacity and coverage. With the exception of some BTS hardware modifications and network software upgrades, EDGE does not necessitate significant hardware changes. \cite{mishra2007advanced}.

\subsection{Network optimization parameters in 2G}
\subsubsection{Key performance indicators in GSM}
Key performance indicators (KPI) are a measure of successful network performance and its
quality. KPIs can be used for the following functions:
\begin{itemize}
    \item [--] To keep track of and improve the performance of the radio network in order to provide better service to subscribers.
    \item [--] To detect unacceptable performance related issues in a cellular network immediately.
    \item [--] To provide detailed information to RF planners.
\end{itemize}

Typically, KPI can be categorized into the following:
\begin{itemize}
    \item \textbf{Accessibility}: Measure the ability to set up a call.
    \item \textbf{Retainability}: Measure the ability to maintain the existing connection.
    \item \textbf{Mobility}: Measure the ability to maintain user connection while moving in network.
    \item \textbf{Integrity}: Measure the degree of service after the service has been accessed by user.
\end{itemize}

\subsubsection{Major KPIs in GSM}
Include all the following parameters:

\begin{itemize}
\item \textbf{Call set up success rate (CCSR)}: Is a measure of successful call attempts over the total number of calls attempts. A call is affirmed to be successful if SDCCH is accessed, occupied then TCH is accessed successfully. CCSR is usually affected due to blocking (SDCCH/TCH) or due to operational issues.
\item \textbf{Dropped call rate (DCR)}: Is a metric that compares the number of dropped calls to the number of calls that were successfully set up. If TCH is dropped from the ager assignment, a call is confirmed to be dropped. From an RF standpoint, DCR is dependent on radio link timeout (RLT), and handover is reduced. RLT is a counter that keeps track of how many SACCH frames have been lost. When one SACCH is lost, it decreases by one, and when one SACCH is received, it increases by two. The RLT value is influenced by the RF environment (interference, low coverage). The probability of an increase in handovers is directly proportional to the probability of an increase in handover failures. These drops can be caused by both RF and operational issues. From an RF standpoint, the neighbor plan should be optimal, frequency clashes in all neighboring cells should be avoided, and all handover control and adjacency parameters should be set to their maximum value. DCR may also rise as a result of operational issues.
\item \textbf{Handover successes rate (HSR)}: Is a metric that compares the number of successful handover attempts to the total number of attempts. Outgoing HSR (out HSR) and incoming HSR (in HSR) are the two types of HSR. Outgoing HSR is dependent on the adjacency and handover control parameters. It can also be affected due to TCH blocking issues in the adjacent cells, leading to blocked handovers. Incoming HSR is also dependent on the adjacency and handover control parameters. It is affected by TCH blocking on the source cell as well.
\item \textbf{Minutes per drop (MPD)}: Is the average number of minutes after which a call gets dropped from the network. For example, if a certain cell has an MPD of 60 minutes, it would imply that a call strays on that cell for 60 minutes on average before getting dropped. MPD is calculated by dividing the minutes of voice traffic with the total of TCH drops.
\item \textbf{TCH raw blocking}: It is a measure of TCH blocking during call set up. It does not include TCH blocking due to incoming handover attempts, or TCH blocking due to incoming directed retry TCH seizure attempts.
\item \textbf{SDCCH blocking}: It is a measure of attempts for SDCCH seizure that could not be facilitated due to non-availability of SDCCH.

\end{itemize}

\subsubsection{TCH drop rate}
It's the difference between the total number of calls attempted and the number of calls dropped. The following are the major causes of the TCH drop rate:
\begin{itemize}
\item \textbf{Low signal strength }: If the signal strength of the uplink (UL) and the downlink (DL) is below the threshold the call gets dropped. To optimize the signal, the output power needs to be checked and the coverage plan.

\textit{The expected action}: Add a signal repeater to increase the coverage area, increase the emission power output, exchange the BTS antenna, add more BTSs to the network, or lock/unlock TRX.

\item \textbf{Bad signal quality}: If the quality of the UL and DL is below the acceptable threshold, the call get dropped caused
by the not sufficient signal quality. To solve this issue interference on the BCCH and TCH, frequency hop (FHOP) configuration needs to be verified.

\textit{The expected action}: Frequency change for BCCG, TCH, change BSIC, mobile allocation index offset (MAIO), FHOP, hopping sequency number (HSN), use a spectrum analyzer to identify the interference source, or resolve the quality problem.

\item \textbf{Sudden connection lost}:
It’s the case were the call gets dropped suddenly. To resolve this problem the BTS’s error logs needs to be consulted.

\textit{The expected action}: Hardware failures needs to be fixed, reset TRX, ensure that the synchronization and the A-bis connection are stable, increase the transmission capacity, or investigate a handover problem.
\item \textbf{Excessive timing advanced (ETA)}: TCH drops caused an excessive TA. To resolve this issue TA max needs to be checked.

\textit{The expected action}: Adjust TA max to a value close to 63 degree, adjust the antenna tilt and reduce its height, or adjust the output power.
\end{itemize}

\subsubsection{TCH congestion}
MSC requests a signaling assignment to the BSC to confirm the MS request for a TCH, if none are available, the BSC refuse the request. The main causer if the TCH congestion is the
lack of the TCH resources. To resolve the TCH congestion, all the TRXs in the cell needs to be checked, verify if the issue is linked to the occurrence of interference or handover drops.

\textit{The expected action}: It uses half of the bandwidth of the full rate codec and operates at 5.6 kbit/s. As a result, network capacity for voice traffic is doubled at the expense of audio quality, or assignment to another cell. Immediately after the detection of an affected SDCCH the call would be transferred to TCH of another adjacent cell.

\subsubsection{SDCCH drop rate}
It is the number of SDCCH channels blocked compared to the number of channels allocated. Causes of SDCCH drop rate are the following:

\begin{itemize}
\item \textbf{Low signal strength UL and DL}: The solutions  is to plan a new site to cover for the low coverage area, increase power output, repair the faulty hardware, or adjust the antenna’s
tilt and height.
\item  \textbf{Poor quality DL and UL}: The expected solution is to isolate the interfering frequency and change it, or do a coverage optimization of either the service cell or the jammer.
\end{itemize}

\subsubsection{SDCCH congestion}
It occurs once the BSC receives an SD request from the MS and all SDCCH resources are
saturated at the time. The causes the SDCCG congestion are the flowing:

\begin{itemize}
\item \textbf{Insufficient system capacity}: The expected solution is to properly plan the location area, configure more SDCCHs, utilize the SDCCH dynamic allocation, or add more TRXs to the system.
\item \textbf{Poor LAC planning}: If the contours of different location areas are at a high traffic area an SDCCH congestion occurs. The expected solution is to resize the location area codes (LAC) to avoided the highly saturated areas, or configure the re-selection condition.
\item \textbf{Interference}: The expected solution is to adjust correctly the minimal threshold of the RACH access, or eliminate the interference by adjusting the frequency.
\end{itemize}

\subsubsection{Handover}

Cause of handover problems:
\begin{itemize}
    \item \textbf{Improper handover settings}: The expected solution is to adjust the cell coverage parameters, adjust the cell reselect offset (CRO), or expand or adjust the TRX configuration between high and low traffic cells.
    \item \textbf{Faulty hardware}: The expected solution to monitor transmission and advisory alarms, check if a transmission disconnection occurred and for faulty advisory cards, or adjust the synchronization.
    \item \textbf{Other causes}: The expected solution is to change the frequencies to avoid signaling failures due to high interference on the TCH, adjust antennas (tilt and height), or rewiring the cables between the TRX and antenna.
\end{itemize}

\section{The third-generation mobile network system}

\subsection{Introduction}
From the second generation, which is entirely digital and includes features such as SMS.
The third generation (3G) of mobile technology was introduced to handle faster data rates and
services such as mobile television, video calls, and so on. 
The UMTS has been significantly enhanced and offers broadband speeds far beyond the original design. These high‐speed enhancements are referred to as high‐speed packet access (HSPA). The totally redesigned
radio access network (AN) in UMTS standard, was the primary enhancement of UMTS in this
initial stage compared to GSM. A new technology known as wide-band code division multiple
access (WCDMA) was invented instead of using Um interface's time and frequency multiplexing capabilities.
The eighth release of the third-generation partnership project (3GPP) specification series
included the first version of LTE. It was able to take advantage of the most up-to-date knowledge and technological developments from HSPA and HSPA+. Consumers can perform tasks on a smart device as efficiently as if they were connected to broadband in their home or office.  In this section, you'll learn about the architecture of UMTS network, including all
of their subsystems and their functions, as well as the various processes that occur within the
network, such as handovers, and the WCDMA concept used to develop the networks. The
architecture enhancements and changes, and most importantly, the network optimization
process that brought UMTS.

\subsection{UMTS Network}
The radio transmission standard for UMTS is the wideband code WCDMA. Many of the
concepts used in GSM have been carried over and improved for UMTS. The SIM card has been
changed into a universal SIM card (USIM). Furthermore, the network has been constructed such
that the GPRS and EDGE advancements may be applied to UMTS. As a result, the needed
investment is maintained to a minimum.
The GSM-inspired core network (CN) topology composed of two user-traffic-dependent
domains which are packet switch (PS) and circuit switch (CS) domains.
For both the GSM and UMTS access networks, the CS and PS domains manage their
relevant traffic types in parallel. All circuit-switched traffic for the UMTS and GSM access networks is handled by the CS domain, which consists of MSC and GMSC entities. Similarly,
the PS domain, which is made up of the GPRS access network, is responsible for all PS traffic
for both GPRS and UMTS, which consist of SGSSN and GGSN entities. In terms of data flow
and connection, the UMTS also has a structure that separates the protocol stack from the
relevant network entities \cite{kukushkin2018introduction}.
For subscriber administration, mobile station roaming and identification, and managing
diverse services, both PS and CS domains utilize the features of the common legacy network
entities such as the HLR in conjunction with the EIR and AuC. As a result, the HLR contains
subscriber information for GSM, GPRS, and UMTS. 

\subsubsection{	UMTS architecture}

With GPRS, the inclusion of packet data necessitated the establishment of more network entities. The UMTS network design was built on the merging of these two network pieces. Figure \ref{umtsArch} depicts the total UMTS architecture with GPRS network. There are three basic parts to the UMTS network design:

\begin{figure}[]
	\centering
	\includegraphics[scale=0.8]{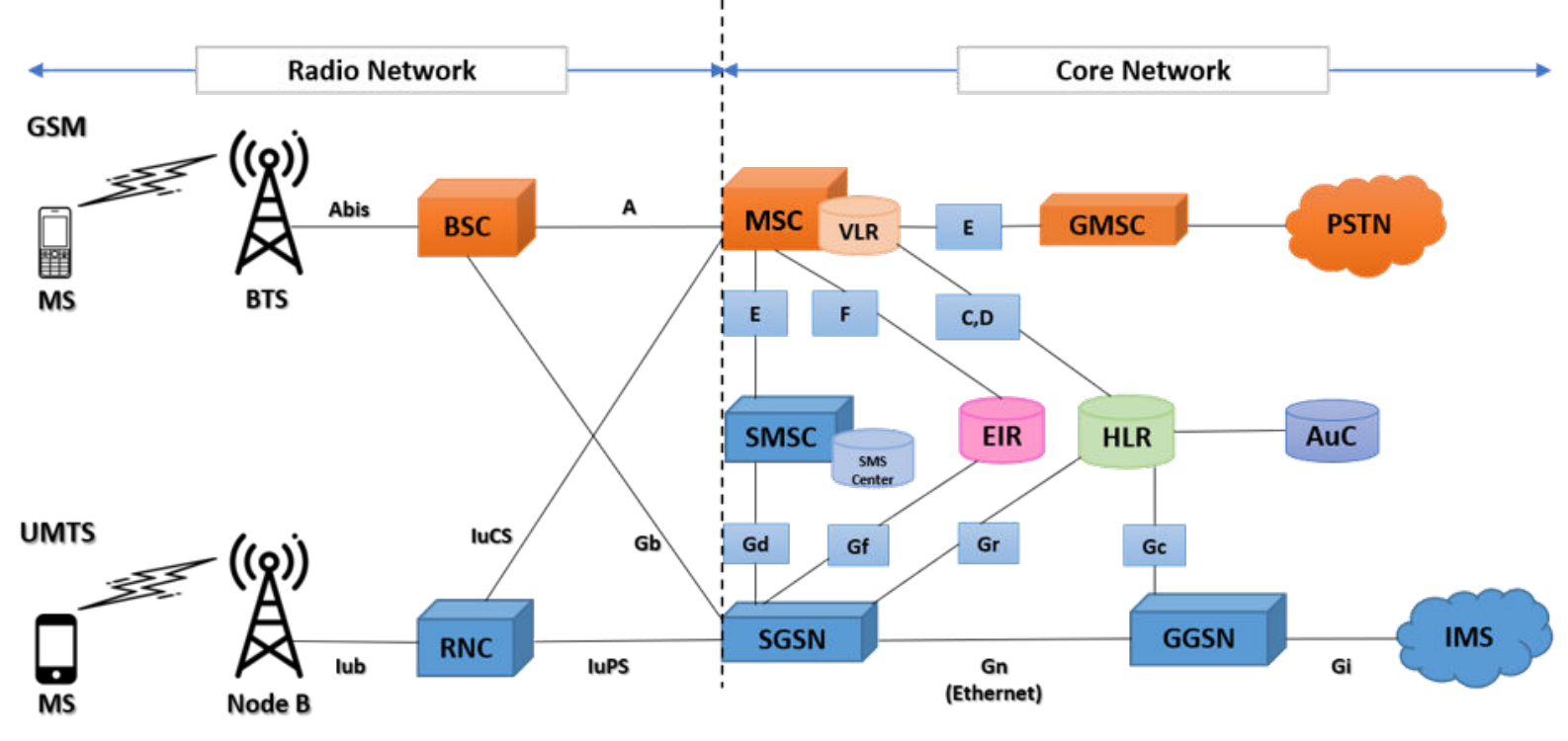}
	\caption{UMTS network architecture connected with legacy network.}
	\label{umtsArch}
\end{figure}

\subsubsection{The radio access network}

\begin{itemize}
    \item  User equipment (UE): The Is a crucial component of the 3G network's overall design. It
serves as the user's last interface. It is made up of:

\begin{itemize}
\item \textbf{RF circuit}: Is in charge of all signal components, both for the receiver and the transmitter. Reduced power consumption was one of the primary problems for the RF power amplifier. WCDMA modulation necessitates the usage of a linear amplifier.
\item \textbf{Base-band processing}: Is a digital circuitry makes up the majority of base-band signal processing.
\item \textbf{Battery}: Must be considered during circuit conception to get a low reduced consumption. Lithium ion (Li-ion) batteries are now commonly used to keep battery is small and light while maintaining their overall life charge.
\item \textbf{USIM}: It is an evolved version of the SIM card employed in GSM, yet it still holds a similar user information, such as MSISDN and IMSI.
\end{itemize}

\item \textbf{UTRAN}: Also known as the radio network subsystem (RNS), is made up of two main components:

\begin{itemize}
\item \textbf{Node-B}: It is the UMTS network's base station, and its primary function is to process signal processing in the Uu interface that performs interleaving, adaptation rate, spreading, channel coding, radio resource management (RRM), and power control \cite{kukushkin2018introduction}.
\item \textbf{Radio network controller (RNC)}: This element manages the radio resources as well as the NodeBs that are connected to it. Call admission control, RRM, code allocation, power control, packet scheduling, handover management, and CS domain encryption are all handled by the RNC.
\end{itemize}
\end{itemize}

\subsubsection{The radio access techniques}
WCDMA, which is a variants of time division duplex (TDD) and frequency division duplex (FDD), were the spread spectrum forms the underlying technique for WCDMA which is the main radio access technology used in UMTS. The following are the basic concepts used in the WCDMA system: channelization, scrambling, channel coding, power control, and handover \cite{kukushkin2018introduction}.

\subsubsection{Spreading}

\begin{figure}[]
	\centering
	\includegraphics[scale=1]{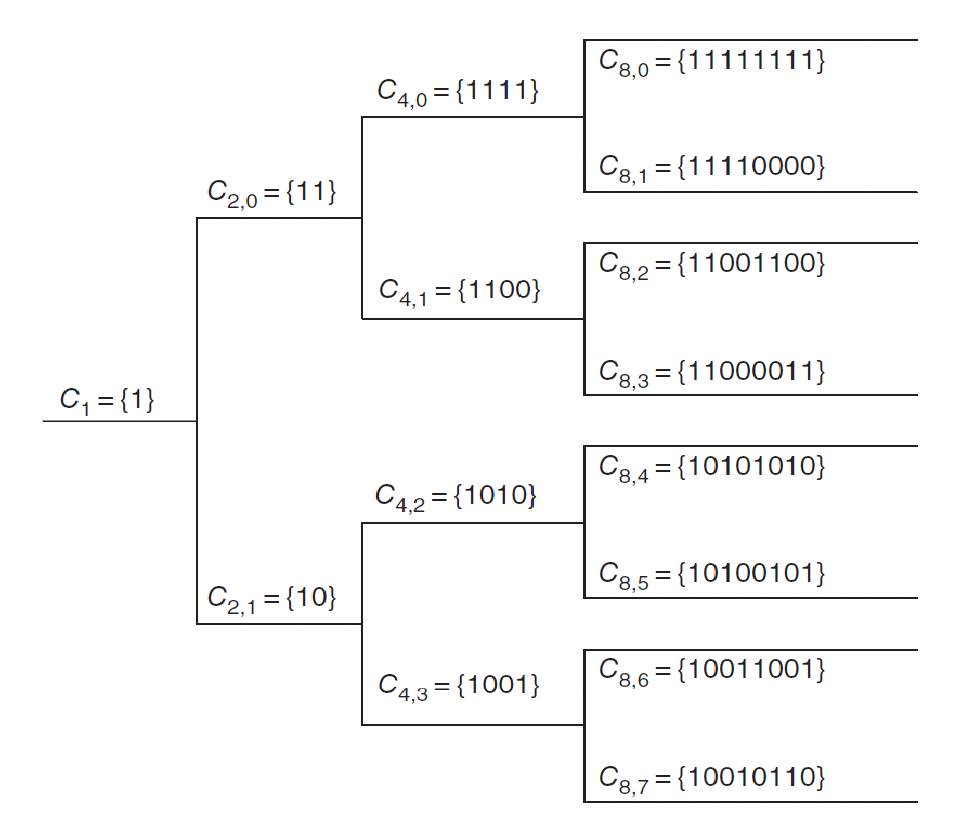}
	\caption{OVSF Code tree.}
	\label{ovsf}
\end{figure}

The spreading technique involves multiplying each user data bit by a four-bit code sequence known as chips (as shown in Figure
\ref{ovsf}). The resulting spread data has the same pseudo noise-like (random appearance) as the
spreading code and is generated at a rate of $4 \times F$. The spread user data signal's occupied spectrum bandwidth widens by a factor of four when the signaling rate is increased by a factor
of four. Di-spreading restores a bandwidth proportional to the signal's intended data rate R. The
spreading sequence is encoded in a particular manner, respecting an orthogonality factor
between data channels. The advantage of using a long orthogonal variable spreading factor
(OVSF) is that it adds redundancy to the transmitted data. The OVSF code length is proportional
to the user's bit rate signal represented in the chip sequence \cite{kukushkin2018introduction}. The principle of OVSF code
generation is depicted in Figure \ref{ovsf}.

\subsubsection{Power control}

The power control is instrumental for WCDMA technique and used basically to:
\begin{itemize}
    \item Calculate the minimum required power for transmission, to get an adequate performance of radio link.
    \item Keep a high signal-to-interference ratio (SIR) throughout the conversation. At three levels of control speed, power control is implemented, namely:
    \begin{itemize}
        \item Open-loop: Both downlink and uplink communications use the open-loop power control mechanism. It provides an initial power setting of the UE for the uplink channel. The signal level calculated by the UE in the down-link channel cannot be used by the FDD WCDMA system.
        \item Open-loop: Both downlink and uplink communications use the open-loop power control mechanism. It provides an initial power setting of the UE for the uplink channel. The signal level calculated by the UE in the down-link channel cannot be used by the FDD WCDMA system \cite{kukushkin2018introduction}.
        \item  Closed-loop: Power control using closed-loop maintains the received signal level of each mobile's uplink signal threshold, reducing interference within the cell.
    \end{itemize}
\end{itemize}

\subsubsection{The handover in UMTS}
There are different types of handover in WCDMA:

\begin{itemize}
\item \textbf{Hard handover}: The term \textit{hard handover} refers to a "hard" change that occurs during subscriber movement
from one area to another. The radio links are ruptured from the last cell to the current cell for hard handover.
\item  \textbf{Soft handover}: The term  \textit{soft handover} happens when the UE is in overlapping coverage area of two cells. soft handover occurs. The UE can establish links to both base stations at the same time, allowing it to communicate with both and combine them using the UE's signal processing module's radio activated key entry (RAKE) receiver capability. The latter process is only possible in downlink; in uplink, only the best signal is chosen. Having multiple links active during the handover process makes the process more reliable and smoother.
\item \textbf{Softer handover}: The term \textit{softer handover} happens when the UE can receive signals from two sectors mastered by one Node B. This can happen because the sectors overlap, or more commonly because of multi-path propagation caused by reflections from buildings, etc. The signals received by Node B, as well as the signals from the two sectors, can be routed to the same RAKE receiver in the uplink, where they can be combined to provide an enhanced signal. The different types of handover in UMTS are illustrated and compared in Figure \ref{3gHO}.
\begin{figure}[]
	\centering
	\includegraphics[scale=0.5]{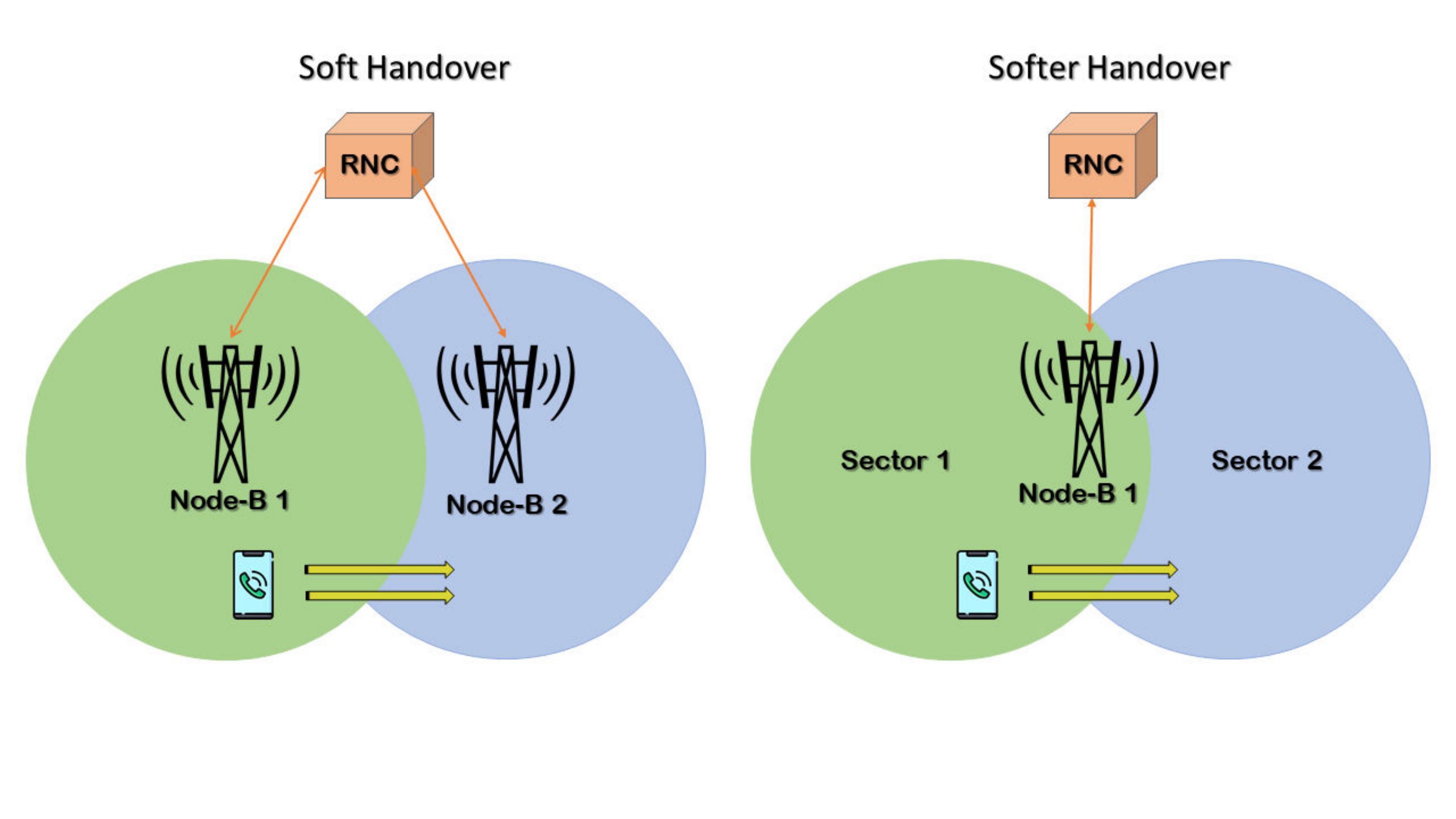}
	\caption{Handover types in UMTS network.}
	\label{3gHO}
\end{figure}

\item \textbf{Inter-RAT handover}:
In many cases, switching from the UMTS radio AN to the 2G network is required. Because they are handing over between different types of RAT, these handovers are referred to
as Inter-RAT ( or I-RAT) handovers. I-RAT handover can be divided into two categories:
\begin{itemize}
\item \textbf{UMTS to GSM handover}: This category of handover is divided into two subcategories:
    
- \textit{Compressed mode handover}: The UE uses the gaps in transmission that occur when using compressed mode handover to analyze the reception of local GSM base stations. \\
- \textit{Blind handover}: This sort of handover happens when the Node B hands off the UE by giving the details of the new cell to it without establishing the link of the mobile for the new cell.

\item \textbf{ GSM to UMTS handover}: The transition from GSM to UMTS occurs to improve performance and is only possible when the conditions are ideal. When this occurs, the UE will be notified by the neighbor list.

\end{itemize}
\end{itemize}

\subsubsection{Cell selection}

The UMTS network's cells can be classified into three different states:

\begin{itemize}
    \item \textbf{Active set}: All of the cells in the active set receive user information. The active set cells in FDD are involved in soft handover. The active set in TDD always consists of one cell. Only active set cells in dynamically variable cell information list will be considered by the UE.
    \item \textbf{Monitored set}: Cells that are not part of the active set but are part of the cell information list.
    \item \textbf{Detected set}: Includes all cells detected by the UE that are neither in the cell information list nor in the active set. Only intra-frequency measurements taken by UEs are eligible for reporting of the detected set's measurements.
\end{itemize}

There are three scenarios in which the state of any given cell can change:
\begin{itemize}
    \item [--] When a cell's signal is strong enough to activate it.
    \item [--] When a cell in the active set becomes weak and needs to be removed.
    \item [--] When the signal of a non-active cell improves to the point where it is superior to that of an active cell, the new stronger cell replaces the weaker one in the active set.
\end{itemize}

\subsection{Major parameters and spectrum allocations in the UMTS network}

The majority of WCDMA spectrum is around 2 GHz, with up-link frequencies ranging from 1920–1980 MHz and down-link frequencies ranging from 2110–2170 MHz. This spectrum is a key component of the IMT-2000 standard, which is used in Europe, Asia and Brazil. Re-farming refers to the process of deploying WCDMA in an existing GSM or CDMA frequency band. The main parameters of WCDMA technology are summarized in Table \ref{umtsPara}.

\begin{table}[]
\centering
	\caption{The main parameters of WCDMA technology \cite{kukushkin2018introduction}.}
	\label{umtsPara}
\begin{tabular}{p{5cm}p{10cm}}
\hline
\textbf{Access method} & \textbf{Direct sequence CDMA}   \\
 \hline
Duplex & FDD/TDD \\
 
Synchronization in radio access & Asynchronous operation \\

Chip rate & 3.84 Mbps \\

Frame length & 10 ms\\

Multi service capability & Multiple services with different QoS in
concurrent operation\\
 
Detection method & Coherent detection using pilot symbols\\
 \hline
\end{tabular}
\end{table}

\subsection{UMTS channels} 

Based on the specifics of the service being offered, any service provided to UE is associated with a radio bearer that defines the setting and configuration of radio link control (RLC),
medium access control (MAC), and the physical layer. At different protocol layers, the information flow associated with a radio access beacon (RAB) is mapped into different types
of channels. The RLC layer creates logical channels, which are then mapped to transport channels in the MAC layer. Control plane and user plane bearers are provided by two types of logical channels:
control channels for signaling and traffic channels for bearer. The existing logical channels are summarized in Table \ref{logicUMTS}.

\begin{table}[]
\centering
	\caption{Most of the existing logical channels in UMTS system.}
	\label{logicUMTS}
\begin{tabular}{p{3cm}p{10cm}p{2cm}}
\hline
\textbf{Channels} & \textbf{Use} & \textbf{Direction}   \\
 \hline
\multicolumn{3}{c}{Control channels} \\
 \hline
Broadcast control channel (BCCH) & Broadcast of the information system message that contains code sequences, cell identifier, timers,...etc. & DL \\

Paging control channel (PCCH) & Incoming calls or other messages should be announced to users in the location area & DL\\
 
Dedicated control channel (DCCH) & Every user with an RRC connection to the RNC has a bidirectional signaling channel. It sends out measurement reports as well as RRC control messages & DL or UL\\
 
Common control channel (CCCH) & Setup of the connection, channel assignment, and cell reselection & DL and UL\\
\hline
\multicolumn{3}{c}{Traffic channels}\\
 \hline
Dedicated traffic channel (DTCH) & Information for a single user's specific service is transferred. Multiple services may be provided to a single user on multiple coexisting DTCHs at the same time & DL and UL\\

 Common traffic channel (CTCH) & SMS cell broadcast message, for example, is a point to multipoint channel that carries information for a group of users & DL\\
 \hline
\multicolumn{3}{c}{Transport channels}\\
 \hline
 
Broadcast channel (BCH) & The BCCH logical channel's transport is provided by this component & DL \\

Paging channel (PCH) &The PCCH logical channel's transport is provided by this component&DL\\
 
Random access channel (RACH) &Before a traffic channel can be assigned, short signaling information must be provided during the initial access to the system&UL\\

Forward access channel (FACH)& A logical channel is carried for a specific UE. RACH's reply& DL\\

Common packet channel (CPCH) &The RACH channel has been extended to allow for the transmission of larger data packets & UL \\

Downlink shared channel (DSCH) &Similar to GPRS, a pool of physical resources is allocated on a time transmission interval (TTI) basis to different users based on packet scheduling policy & DL\\
\hline
\end{tabular}
\end{table}

\subsection{RRC states}

The RRC state machine has two modes: RRC idle mode and RRC connected mode.
\subsubsection{RRC idle mode} 
The UE, in idle mode, is identified by non-access stratum (NAS) identities such as TMSI and IMSI rather than by UTRAN. The network can page the UE based on the terminal's position in the registration area. As long as the mobile remains in the same registration area, it is not necessary for the terminal to initiate an RRC connection. When the mobile enters a new registration area, it must switch to RRC connected mode to make a location update \cite{kukushkin2018introduction}.

\subsubsection{RRC connected mode}
\begin{figure}[]
	\centering
	\includegraphics[scale=1]{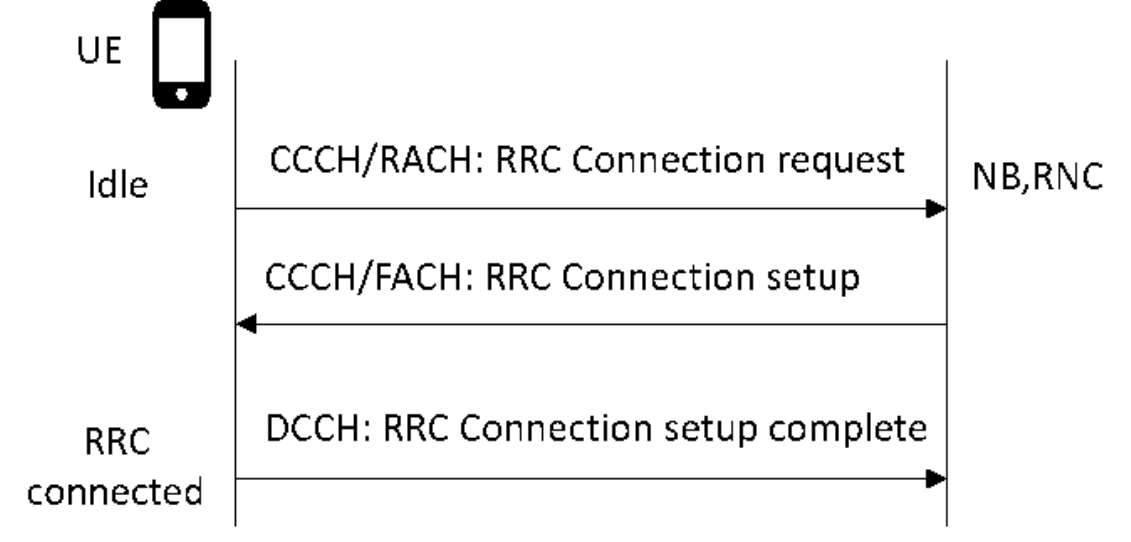}
	\caption{RRC connection establishment.}
	\label{rrcConect}
\end{figure}

The UE can transmit data using allocated radio resources in RRC connected mode. So, each UE has a temporary radio network identity (RNTI) that identifies the mobile phone, in common channels such as RACH or CPCH. Figure \ref{rrcConect} depicts the message flow for establishing an RRC connection. The uplink CCCH mapped to RACH, the UE sends an RRC connection request message. The UTRAN, like any other request from NAS, must establish a signaling radio bearer (SRB) between the UTRAN and the UE. The mobile is identified by its initial \textit{UE ID},  which is a NAS identifier similar to the IMSI or TMSI. The RRC connection request content includes the initial UE ID \cite{kukushkin2018introduction}.

\subsection{UMTS network optimization}
\subsubsection{KPIs in UMTS}
The KPIs of the UMTS network can be divided into the following subcategories:
\begin{itemize}
    \item \textbf{Accessibility}: It is a user's ability to successfully connect to the UMTS network and use the available services.
    \item \textbf{Retainability}: It is defined as a user's ability to keep a requested service for the specified amount of time.
    \item \textbf{Mobility}: It refers to a user's ability to move between neighboring UMTS and GSM cells while maintaining their requested service (PS or CS).
    \item \textbf{Integrity}: represent the user's perception of quality.
\end{itemize}
	
CS or PS call drop, RRC connection setup failure, HO failure (3G-3G), IRAT HO failure (3G-2G), and power congestion issue are the major UMTS network KPIs.

\subsubsection{KPIs optimization }
All of the KPIs mentioned above aid in the continuous optimization of the network, which is an essential part of its operations. By monitoring them, several issues within the network can be identified and resolved as needed, as detailed in tables \ref{csCalldrop}, tables \ref{psCalldrop}, table \ref{rrcConnect}, table \ref{hoFailer}, table \ref{IRAT}, and table \ref{power}.

\begin{table}[]
\centering
	\caption{CS call drop optimization.}
	\label{csCalldrop}
\begin{tabular}{p{5cm}p{10cm}}
\hline
\textbf{Possible reasons} & \textbf{Resolutions}   \\
 \hline
Poor coverage & Adjust the antenna’s tilt and azimuth. \\

Neighbor planning issues & Check for neighbors planning of the cells in the area \\

Received total wideband power (RTWP) issue & Tracing the source of the external interference by using a spectrum analyzer or check the related hardware \\

Alarms & UTRAN rectification \\

Overshooting issue & RF conditions needs to be checked \\

Scrambling code planning (SCP)  clash (cells coverage) & Re-plan the Pluripotent stem cells (PSC) \\
 \hline
\end{tabular}
\end{table}

\begin{table}[]
\centering
	\caption{PS call drop optimization.}
	\label{psCalldrop}
\begin{tabular}{p{5cm}p{10cm}}
\hline
\textbf{Possible reasons} & \textbf{Resolutions}   \\
 \hline
Poor RF conditions & Antenna adjustments \\

Missing neighbor issue & Check for neighbors planning of the cells in the area \\

RTWP issue & Tracing the source of the external interference by using a spectrum analyzer or the hardware needs to be checked \\

Alarms & UTRAN rectification \\

Tracing the particular UE & Check the call history records (CHR) logs\\
 \hline
\end{tabular}
\end{table}

\begin{table}[]
\centering
	\caption{RRC connection failure optimization.}
	\label{rrcConnect}
\begin{tabular}{p{5cm}p{10cm}}
\hline
\textbf{Possible reasons} & \textbf{Resolutions}   \\
 \hline
RRC failure & Apply a RRC reject analysis \\

Power congestion & Increase the power output or check for RTWP issues \\

CE congestion & Load balancing Code congestion Check the HSDPA codes \\

No reply & UE issue \\

UE issue & Cell scenario analysis \\
 \hline
\end{tabular}
\end{table}

\begin{table}[]
\centering
	\caption{HO failure optimization.}
	\label{hoFailer}
\begin{tabular}{p{5cm}p{10cm}}
\hline
\textbf{Possible reasons} & \textbf{Resolutions}   \\
\hline
Overshooting issue & Check RF conditions \\

Missing neighbor issue & Check cells planning in the area \\

 Handover event threshold & Retune handover event threshold \\

 PSC clash & Replan the cells PSC \\

 Poor coverage & Antenna adjustments \\
 \hline
\end{tabular}
\end{table}

\begin{table}[]
\centering
	\caption{IRAT HO failure optimization.}
	\label{IRAT}
\begin{tabular}{p{5cm}p{10cm}}
\hline
\textbf{Possible reasons} & \textbf{Resolutions}   \\
\hline
GSM cell discrepancies & Remove all GSM cell discrepancies  \\

Interference on the GSM cell & Trace and resolve the interferences\\

IRAT parameters & Retune IRAT parameters \\

Particular cell issue & Filter the cells with continuous IRAT failures\\
 \hline
\end{tabular}
\end{table}

\begin{table}[]
\centering
	\caption{Power congestion optimization.}
	\label{power}
\begin{tabular}{p{5cm}p{10cm}}
\hline
\textbf{Possible reasons} & \textbf{Resolutions}   \\
\hline
RTWP issue & Check hardware \\
Power output issue & Increase TCP and CPICH power to increase
coverage and capacity\\
Overshooting cell & Check RF conditions \\
 \hline
\end{tabular}
\end{table}

\section{The fourth-generation mobile network system}
\subsection{Overview of the LTE network}

LTE's main goal is to provide a higher data rate than UMTS, as well as low latency and packet-optimized radio access technology that allows flexible bandwidth deployments. At the
same time, its network architecture was created with the goal of supporting packet-switched traffic while maintaining high quality of service and mobility. When compared to previous
cellular systems, LTE has introduced many new technologies. These latter allow LTE to operate more efficiently in terms of spectrum utilization while also delivering higher data rates that are requires:
\begin{itemize}
    \item \textbf{OFDM}: It is the signal format for LTE, it allows a high data bandwidth to be transmitted efficiently while remaining
resistant to interference and reflections. Since the traffic was carried on a several number
of carriers, the system was able to cope even if some were lost due to interference from
reflections, the uplink and downlink had different access schemes.
\item \textbf{Orthogonal frequency division multiple access (OFDMA)}: It has many advantages
including its robustness in the face of multipath fading and interference. It is used in the
downlink, but in the uplink, a single carrier frequency division multiple access (SCFDMA) instead because the peak of average power ratio of SC-FDMA is lower than that
of OFDMA, which will extend battery life.
\item \textbf{Multiple input multiple output (MIMO)}: The multi-path signals could be used to
advantage and increase throughput by using MIMO. When using MIMO, multiple
antennas are required to distinguish between the different paths.
\item \textbf{System architecture evolution (SAE)}: While the term "LTE" refers to the evolution of
UMTS radio access via the evolved UTRAN (E-UTRAN), the term "SAE" refers to the
evolution of non-radio aspects such as the evolved packet core (EPC) network. The
evolved packet system (EPS) is made up of LTE and SAE. To route IP traffic from a
gateway in the packet data networks (PDN) to the UE, EPS employs the concept of EPS
bearers. Between the gateway and the UE, a bearer is an IP packet flow with a defined
quality of service (QoS). Together, the E-UTRAN and EPC set up and release bearers as
needed by applications.
\item \textbf{IP data}: LTE is a completely IP-based data system. Circuit switched voice was
included in UMTS, but it is not included in LTE. Originally, it was thought that operators
would provide data and that voice would be delivered via over-the-top (OTT)
applications. To address this, the GSM-Association (GSMA) established the voice over
LTE (VoLTE) scheme as the industry standard. The construction of an IP multimedia
subsystem (IMS) core was necessary for VoLTE, but it cost delayed the rollout of this
functionality.
\end{itemize}

\subsection{LTE network architecture}
\begin{figure}[]
	\centering
	\includegraphics[scale=0.9]{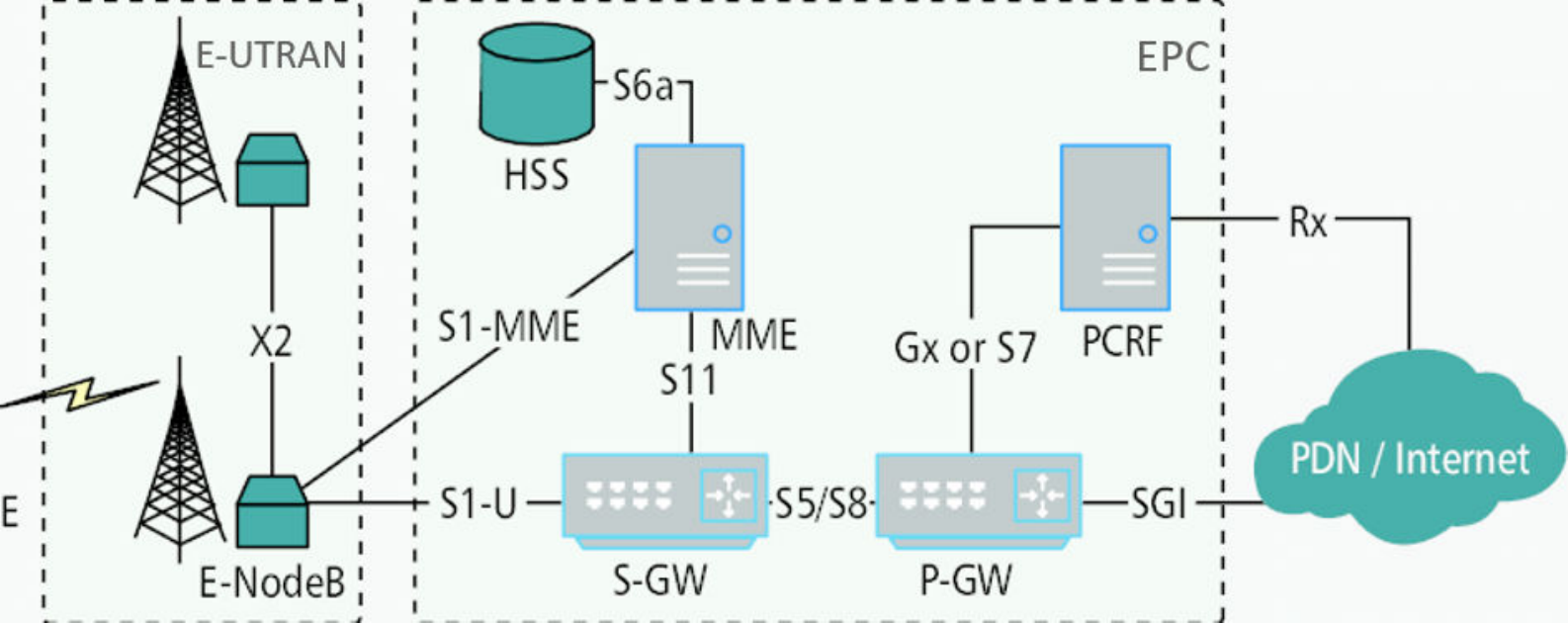}
	\caption{LTE Network architecture.}
	\label{lteArch}
\end{figure}

LTE was created to support only PS services, with the goal of providing seamless IP
connectivity between the UE and the PDN without causing any interruption to the end users'
services when moving. 
The uper-level of LTE network is consisted of following three main entities (Figure \ref{lteArch}):
The EPC, the UE, and the E-UTRAN.

\subsection{Radio access network in LTE}

\subsubsection{The UE}

The internal architecture of LTE user equipment is similar to that of UMTS. The
following important modules were included in the UE:

\begin{itemize}
    \item \textbf{Mobile termination (MT)}: It manages all aspects of communication.
    \item \textbf{Terminal equipment (TE)}: It disconnects the communication stream.
    \item \textbf{Universal integrated circuit card (UICC)}: For LTE equipment, this is referred to the USIM card. This latter is similar to a 3G SIM card in that it stores user-specific data, such as the user's phone number, home network identity and security keys, ...etc.
\end{itemize}

\subsubsection{The E-UTRAN}

The eNodeB or eNB, are the single component of the E-UTRAN, and they manage radio
communications between the UE and the EPC. Each eNB controls mobile devices as a base
station for one or more cells. The base station that communicates with the mobile is known as
the serving eNB. LTE Mobile only connects with one base station and one cell at a time, and
the eNB provides two basic roles:
\begin{itemize}
    \item Using the LTE air interface's digital signal and analogue processing functionality, the
eNB sends and receives radio transmissions to all mobiles.
\item The eNB manages all the low-level operations of the mobiles by sending them signaling
messages such as handover commands.
\end{itemize}

The S1 interface connects each EPC to the eNB, and the X2 interface connects it to nearby
base stations, which is primarily used for signaling and packet forwarding during handover.

\subsubsection{LTE core network}
It is made up of components that perform tasks such as mobility authentication, management,
quality of service, download and upload IP packet routing, and IP address allocation, among
others. The EPC's flat IP architecture enables the network to efficiently and cost-effectively
handle large amounts of data traffic, and it's made up of:
\begin{itemize}
\item \textbf{Mobility management entity (MME)}: It manages all signaling between the UEs
and the EPC, as well as between the eNodeBs and the EPC. The MME's signaling is
also known as NAS signaling because it is done using the NAS protocol. Through the
S1-AP interface, the MME connects to the eNodeB and performs authentication. It
establishes a connection with the HSS and orders the authentication information for
the user attempting to join the network. The functions of the MME are as follows:

\begin{itemize}
\item Authentication: After exchange the authentication information between the HSS
and the UE, UEs can authenticate to the network.
\item Mobility management: Allows subscribers to move around the network or
between networks.
\item  Location update: tracks the subscriber's current location and state within the
network.
\item  Bearer establishment: If more gateways are available, bearers are established by
determining on a gateway routes to the Internet.
\item  Handover support: Enables inter-eNodeB handover (for handover on the S1
interface).
\item  While the eNodeB has handover capabilities, when the X2 interface is
unavailable, the MME sends handover messages between eNodeBs.
\end{itemize}
The MME is also in charge of generating and allocating temporary identities to
UEs, as well as terminating NAS signaling. It verifies the UE's permission to camp
on the PLMN of the service provider and enforces UE roaming restrictions. The
MME is the network's ciphering/integrity protection and security key management
termination point for NAS signaling. It also advocates for the legal interception of
signaling. With the S3 interface terminating at the MME from the SGSN, it provides
the control plane function for mobility between 2G/3G access and LTE ANs. For
roaming UEs, the MME also terminates the S6a interface to the home HSS.

\item \textbf{Home subscriber server (HSS)}: Stores the SAE subscription of the user,
including the EPS-subscribed QoS profile and any roaming access restrictions. It also
contains data on the PDNs to which the user has access. An access point name (APN)
or a PDN address could be used. Furthermore, the HSS stores dynamic data such as
the MME's identity to which the user is currently connected or registered. The HSS
can also include the AuC, which generates authentication and security key vectors.

\item \textbf{Serving gateway (SGW)}: When the UE moves between eNodeBs, the S-GW
serves as the local mobility anchor for the data bearers, transferring all user IP packets.
When the UE is idle, it saves information about the bearers and temporarily buffers
downlink data while the MME initiates UE paging to re-establish the bearers. The SGW collects data for legal interception and charging. It also acts as a mobility anchor
for other technologies like UMTS and GPRS, allowing them to communicate with
each other.

\item \textbf{Packet data network gateway (P-GW)}: Is in charge of assigning IP
addresses to UEs, charging and QoS enforcement based on PCRF rules. The P-GW is in charge of filtering downlink user IP packets and distributing them to the various
QoS-based bearers. This is done using traffic flow templates (TFTs). For guaranteed
bit rate (GBR) bearers, the P-GW performs QoS enforcement. It also acts as a mobility
anchor for non-3GPP technologies like CDMA2000 and WiMAX networks, allowing
them to communicate with each other.

\item \textbf{Policy and charging rules function (PCRF)}: Is housed in the P-GW, it is responsible for policy control
decision-making as well as controlling the flow-based charging functionalities. The
PCRF provides the QoS authorization that determines how a particular traffic data
will be handled in the PCEF and guarantees that it is consistent with the subscription
profile of the user.

\end{itemize}

\subsection{LTE Interfaces}

\begin{itemize}
\item \textbf{LTE-Uu}: It connects the UEs to the eNodeBs. It is responsible for all signaling messages between the MME and the eNodeB, and the data traffic between the S-GW and UE.
\item \textbf{S1}: For both the user and control planes, the S1 interface connects the E-UTRAN
and the EPC. It is divided into two parts: the S1-AP (GTP-AP) which belongs to the
control plane, and the S1-U (GTP-U), which belongs to the user plane.
\item \textbf{X2}: Allows two or more eNodeBs to interconnect with each other. The X2 interface
is divided into two parts: the X2-C which connects the control planes of eNodeBs,
and the X2-U, which connects the user planes of eNodeBs. The structure of the X2-
C and X2-U interfaces is identical to that of the S1 interface.
\item \textbf{S5}: Connects S-GW and P-GW, it provides user-plane tunneling and tunnel
management. It is based on the GTP protocol and is used to connect a non-collocated
P-GW for PDN connectivity, and S-GW relocation due to UE mobility \cite{ahmadi2013lte}.
\item \textbf{SGi}: It is responsible for connecting the P-GW to the packet data network. As a
provision for IMS services, a packet data network can be an operator's external public
or private packet data network or an intra-operator packet data network \cite{ahmadi2013lte}.
\end{itemize}

\begin{figure}[]
	\centering
	\includegraphics[scale=1]{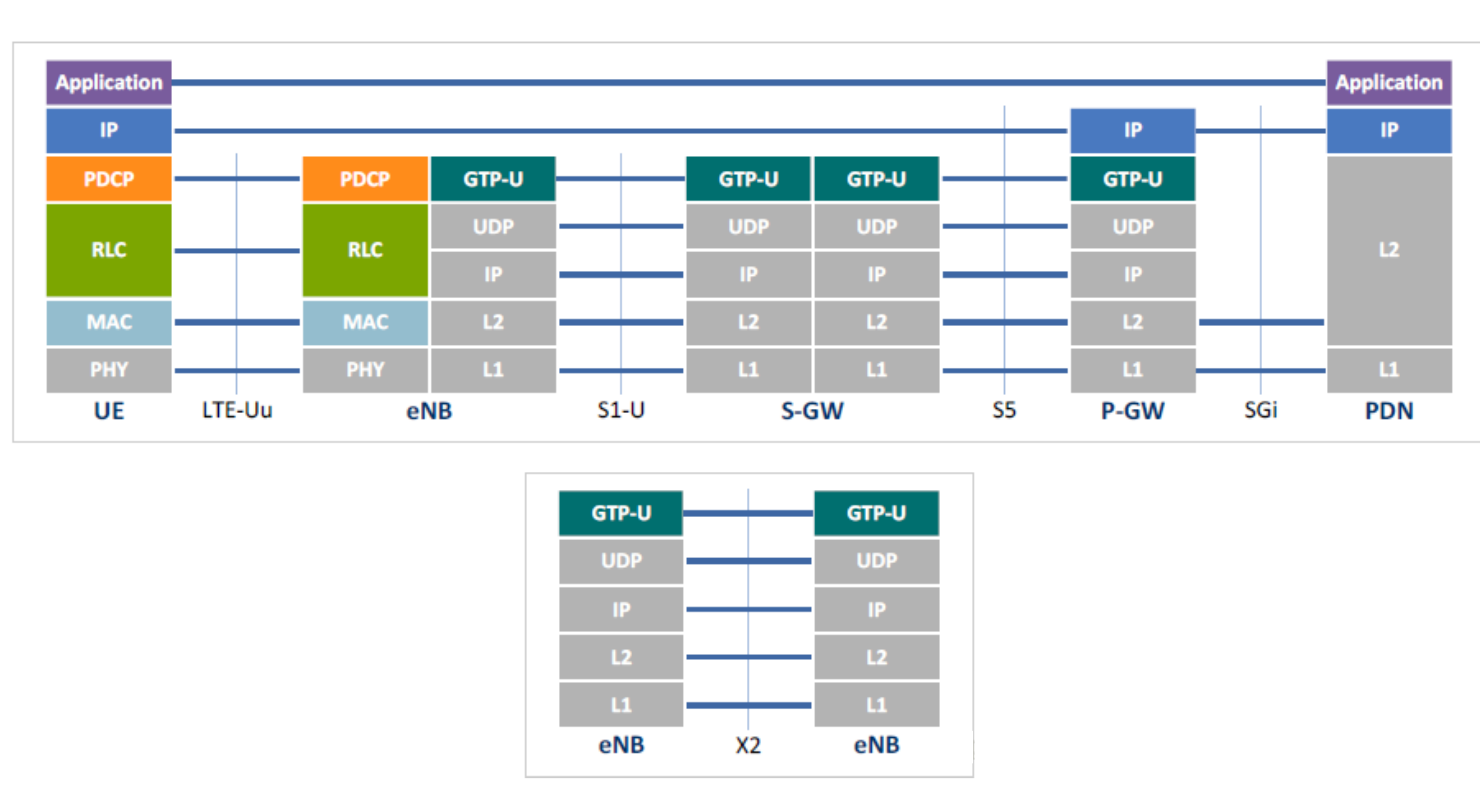}
	\caption{LTE user plane protocol stacks.}
	\label{proto}
\end{figure}

Figure \ref{proto} shows the user plane protocol stacks for the LTE network.

\subsection{Handover in LTE}

When it comes to mobility, such as handover and re-selection, LTE faces significant
challenges. In comparison to UMTS, LTE technology makes handover more difficult. Intra-LTE handovers, as well as handovers between LTE and UTRAN, LTE and GSM-EDGE radio
access network (GERAN), and other mobile networks, are all handled by LTE.
\subsubsection{Inter-RAT mobility} 
\begin{figure}[]
	\centering
	\includegraphics[scale=0.6]{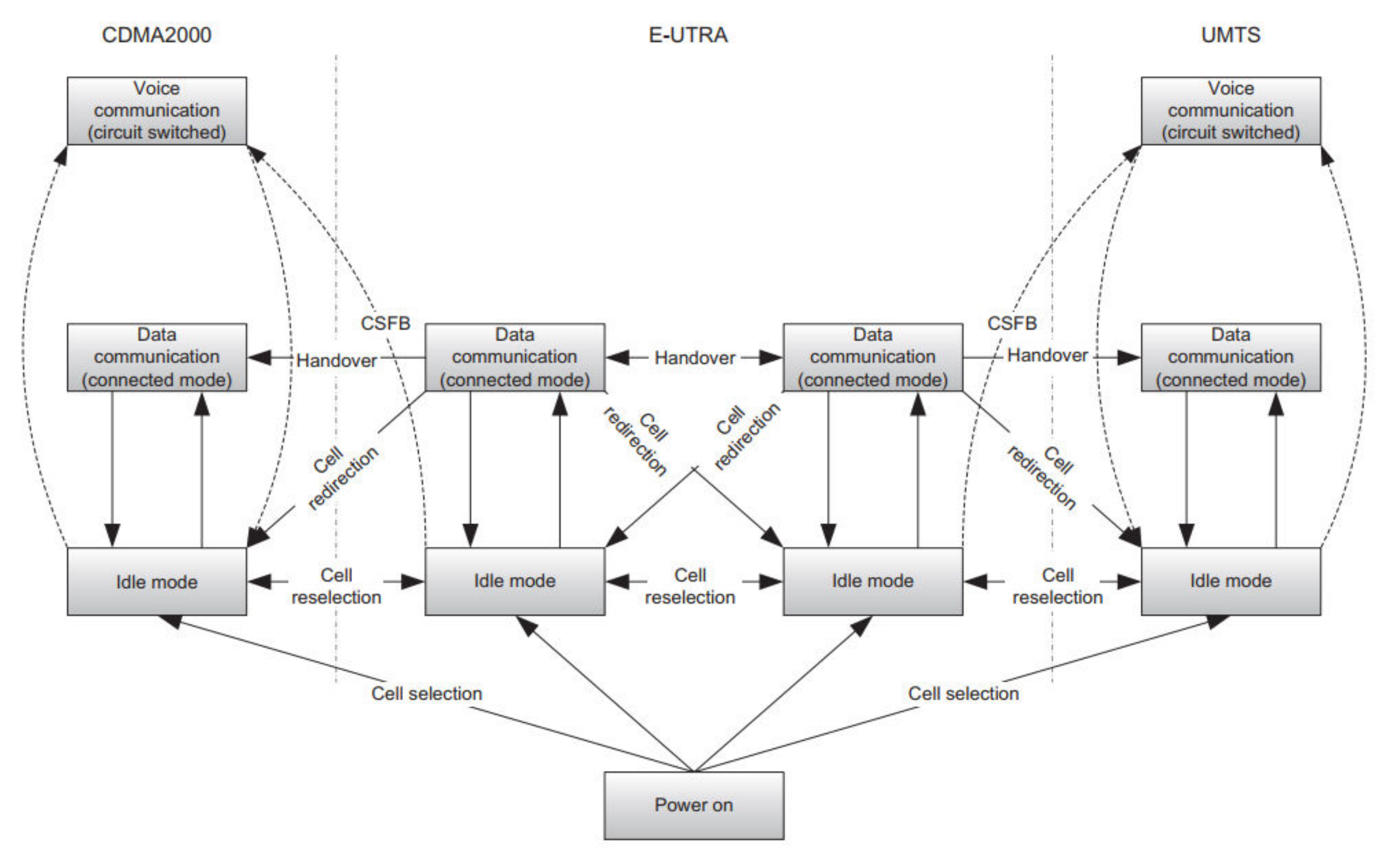}
	\caption{Inter-RAT mobility procedures.}
	\label{IRATLTE}
\end{figure}

When a UE is turned on, it conducts a cell search to locate and chose an appropriate cell to
lodge on. Cell reselection is an idle mode state and the mechanism for switching cells after a
UE has been camped on one \cite{sesia2011lte}. Handover is a connected mode state that occurs when two
connected states of the different or the same radio access technologies are switched over.
Redirection is a process that changes the state of the UE from connected to idle, often for circuit
switch full back (CSFB) purposes.
Because a handover is not required and may not be supported by the network or the UE in
the case of CSFB, redirection is the default technique to use. Moreover, due to factors like interRAT measurement cell/delay detection, 10 time-to-trigger, network handover initiation, and
where the call originated or terminated, redirection can happen before handover in some cases
\cite{sesia2011lte}. LTE signaling and state transitions are depicted in Figure \ref{IRATLTE}.

\subsubsection{Mobility towards home eNodeBs}
Many functions from the source LTE RAN node and the MME are required for mobility
towards home eNodeBs (HeNBs). The source RAN node should include the closed subscriber
group identity (CSG ID) and the access mode of the target HeNB in the handover required
message to the MME in addition to the E-UTRAN cell global identifier (ECGI) so that the
MME can execute access control to that HeNB. When MME rejects the handover, it sends back
handover preparation failure message if MME fails the access control and the target HeNB is
in closed access mode.

\subsubsection{Seamless handover}
If the source eNodeB selects seamless handover mode for a given bearer, it suggests to the
target eNodeB that a GTP tunnel must be established to operate the downlink data forwarding.
\subsubsection{ Lossless handover}
If the source eNodeB selects the lossless mode for a given bearer, those user plane downlink
packets that have been processed by the packet data convergence protocol (PDCP) are still
locally buffered because it’s have not been neither sent nor acknowledged by the UE but it will
be forwarded over the X2 interface.

\subsection{LTE radio access type}

On both the FDD and TDD fields, LTE is being installed around the world (Table \ref{tabLTE}). In
FDD mode, two frequency channels with a guard band are used for communication, while in
TDD mode, a single frequency is employed to transmit and receive data using different
timeslots. FDD employs a paired spectrum, whereas TDD employs an unpaired spectrum \cite{mishra2018fundamentals}.
Table \ref{tabLTE} summarizes the differences between LTE-FDD and LTE-TDD.

\begin{table}[]
\centering
	\caption{The difference between LTE-FDD and LTE-TDD \cite{kibret2014study}}
	\label{tabLTE}
\begin{tabular}{p{4cm}p{5.5cm}p{5.5cm}}
\hline
Parameter & \textbf{FDD}  & \textbf{TDD} \\
 \hline
Spectrum & Paired spectrum & Unpaired spectrum \\
Traffic & The asymmetry depends
on available spectrum &
The asymmetry is dynamically
adjustable \\
Frequency/Time separation & Need a guard band & Need a guard period \\
Interferences intra-system & Unexpected to occur & eNodeBs synchronization is required \\
Size of cell & Fit large or small cells & Fit small cells because of guard period \\
Costs of hardware & Complicate duplexer
causes a high cost & There is no significant cost \\
 \hline
\end{tabular}
\end{table}

\begin{figure}[]
	\centering
	\includegraphics[scale=0.6]{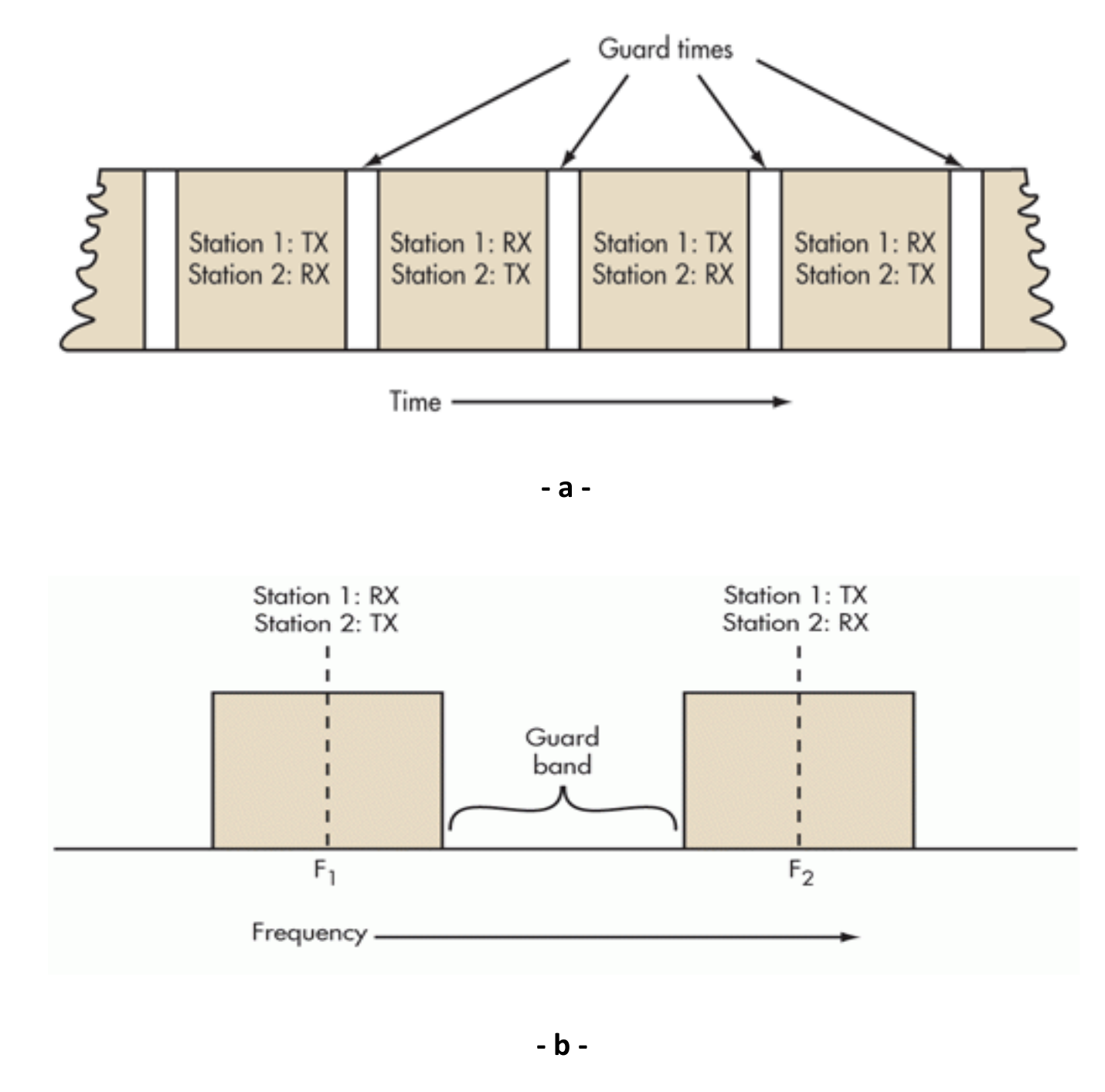}
	\caption{LTE radio access type. (a) TDD mode, (b) FDD mode.}
	\label{tddfdd}
\end{figure}

Figures \ref{tddfdd} (a) and (b) depict in details the principle of TDD and the FDD modes.

\subsection{LTE optimization and KPI analysis}

Typically, KPI in LTE can be splitted into many subcategories:
\begin{itemize}
    \item \textbf{Accessibility}: Used to determine if the users' requested services can be accessed
in a specific condition, as well as the quality of availability when users require it.
For example, a user may request access to the network, a voice call, or exchange, ...etc.

\item \textbf{Retainability}: Used to determine how well a network can keep a user's possession
or provide services to them.

\item \textbf{Mobility}: Used to assess the performance of a network's ability to handle user
movement while maintaining service to the user.

\item \textbf{Integrity}: Used to assess a network's character or honesty to its users, such as
throughput, latency, and the number of users served.

\item \textbf{Availability}: Used to determine whether a network is available, ready, or suitable
for users to use services.

\item \textbf{Utilization}: Used to determine whether a network's capacity has been reached
and whether its resource has been depleted.

\end{itemize}

Table \ref{lteKPI} summarizes in details the LTE KPI and its indicators.

\begin{table}[]
\centering
	\caption{KPIs of LTE RAN.}
	\label{lteKPI}
\begin{tabular}{p{5cm}p{8cm}}
\hline
\textbf{LTE KPI} & \textbf{Indicators}   \\
\hline
Accessibility &  RRC setup success rate.\\
&  ERAB setup success rate. \\
&  Call setup success rate.  \\

Retainability &  Call drop rate. \\
&  Service call drop rate. \\

Mobility &  Intra-frequency handover out success rate.\\
&  Inter-frequency handover out success rate. \\
&  Inter-RAT handover out success rate (LTE to WCDMA). \\

Integrity &  E-UTRAN IP throughput. \\
&  IP Throughput in DL. \\
&  E-UTRAN IP latency. \\
Availability &  E-UTRAN cell availability. \\
Utilization & Mean active dedicated EPS bearer utilization.\\
 \hline
\end{tabular}
\end{table}

\section{Conclusion}

This document describes the main aspects of a GSM network, it’s architecture with all the different subsystems and entities, all the releases of the 2G network (GPRS, EDGE) with the
upgrades that were made to the network. Also defined some of the network key functions such
as the handover and roaming. Moreover, the procedures and methods that can be utilized to
optimize mainly coverage and capacity of the cellular network. It has been shown that the
optimization process requires a network analysis with the use of different types of KPIs and
troubleshooting and fine-tuning parameters to maintain the reliability and quality of the
network.

Besides, this document described the main aspects of the UMTS and LTE networks, its architecture
with all the different subsystems and entities, all the enhancements and changes that were made
to the network compared to the GSM. Also, some of the network’s key functions was defined
such as all the different types of handovers and all the new functions and concepts that the
WDCMA brought, all the logical channels that are used for the communication between the
different entities. KPIs optimization and suggested solution are also detailed for all the discussed technologies.

\bibliographystyle{unsrtnat}
\bibliography{template}  






\end{document}